\documentclass[12pt]{amsart}
\usepackage{hyperref}
\usepackage{amssymb}
\usepackage{amsthm}
\usepackage{latexsym}
\usepackage{epic, eepic, graphicx}
\usepackage{graphicx} \DeclareGraphicsExtensions{.ps}
\theoremstyle{plain}
\newtheorem{thm}{Theorem}
\newtheorem{prop}{Proposition}[section]

\theoremstyle{definition}

\newtheorem{rem}{Remark}[section]

\newtheorem{defn}[prop]{Definition}

\newcommand{\nwc}{\newcommand}

\newcommand{\be}{\begin{equation}}
\newcommand{\ee}{\end{equation}}

\nwc{\IA}{\mathbb{A}} 
\nwc{\IB}{\mathbb{B}} 
\nwc{\IC}{\mathbb{C}} 
\nwc{\ID}{\mathbb{D}} 
\nwc{\IE}{\mathbb{E}} 
\nwc{\IF}{\mathbb{F}} 
\nwc{\IG}{\mathbb{G}} 
\nwc{\IH}{\mathbb{H}} 
\nwc{\IN}{\mathbb{N}} 
\nwc{\IP}{\mathbb{P}} 
\nwc{\IQ}{\mathbb{Q}} 
\nwc{\IR}{\mathbb{R}} 
\nwc{\RR}{\mathbb{R}} 
\nwc{\IS}{\mathbb{S}} 
\nwc{\IT}{\mathbb{T}} 
\nwc{\TT}{\mathbb{T}} 
\nwc{\IZ}{\mathbb{Z}} 
\def\bbbone{{\mathchoice {1\mskip-4mu {\rm{l}}} {1\mskip-4mu {\rm{l}}}
{ 1\mskip-4.5mu {\rm{l}}} { 1\mskip-5mu {\rm{l}}}}}

\nwc{\tM}{\widetilde{M}}
\nwc{\tOmega}{\widetilde{\Omega}}

\nwc{\cD}{\mathcal{D}}
\nwc{\cH}{\mathcal{H}}
\nwc{\cK}{\mathcal{K}}
\nwc{\cL}{\mathcal{L}}
\nwc{\cM}{\mathcal{M}}
\nwc{\cO}{\mathcal{O}}
\nwc{\cP}{\mathcal{P}}
\nwc{\cR}{\mathcal{R}}

\nwc{\bSigma}{\boldsymbol{\Sigma}}
\nwc{\balpha}{\boldsymbol{\alpha}}

\nwc{\eps}{\epsilon}
\nwc{\vareps}{\varepsilon}
\nwc{\bep}{\boldsymbol{\epsilon}}
\nwc{\tkappa}{\tilde\kappa}

\nwc{\tr}{\operatorname{tr}}
\nwc{\rest}{\restriction}
\nwc{\trap}{K}
\nwc{\supp}{\operatorname{supp}}
\nwc{\rk}{\operatorname{rank}}
\nwc{\Res}{\operatorname{Res}}
\nwc{\Spec}{\operatorname{Spec}}
\nwc{\Vol}{\operatorname{Vol}}
\nwc{\Var}{\operatorname{Var}}
\nwc{\Op}{\operatorname{Op}_\hbar}
\nwc{\defeq}{\stackrel{\rm{def}}{=}}
\nwc{\la}{\langle}
\nwc{\ra}{\rangle}

\renewcommand{\Re}{\operatorname{Re}}
\renewcommand{\Im}{\operatorname{Im}}

\title{Spectral problems in open quantum chaos}
\author{St\'ephane Nonnenmacher}
\address{Institut de Physique Th\'eorique, 
CEA/DSM/IPhT, Unit\'e de recherche associ\'ee au CNRS,
CEA/Saclay, 91191 Gif-sur-Yvette, France}
\email{snonnenmacher@cea.fr}

\begin{document}
\begin{abstract}
We present an overview of mathematical results and methods relevant for
the spectral study of semiclassical Schr\"odinger (or wave) operators
of scattering systems, in cases where the corresponding classical
dynamics is chaotic; more precisely, we assume that in some energy
range, the classical Hamiltonian flow admits a
fractal set of trapped trajectories, which hosts a 
chaotic (hyperbolic) dynamics. The aim is then to connect the information on this
trapped set, with the distribution of
quantum resonances in the semiclassical limit. 

Our study encompasses
several models sharing these dynamical characteristics: 
free motion outside a union of convex hard obstacles, scattering by
certain families of compactly supported potentials, geometric
scattering on manifolds with (constant or variable) negative
curvature. We also consider the toy model of open quantum maps, and
sketch the construction of
quantum monodromy operators associated with a Poincar\'e section for a
scattering flow.

The semiclassical density of long living resonances exhibits a
fractal Weyl law, related with the fact that the corresponding
metastable states are ``supported'' by the fractal trapped set (and its
outgoing tail). We also
describe a classical condition for the presence of a gap in the resonance spectrum,
equivalently a uniform lower bound on the quantum decay rates, and
present a proof of this gap in a rather general situation, using
quantum monodromy operators.

\end{abstract}

\maketitle
\tableofcontents

\section{Introduction}

This review article will present some recent results and methods in the study
of 1-particle quantum or wave scattering systems, in the
semiclassical/high frequency limit,
in cases where the corresponding classical/ray dynamics is
chaotic. 

The study of such systems has a long history in
physics and mathematics, ranging from mesoscopic semiconductor physics to number theory.
We will focus on some mathematical aspects, adopting a  {\it quantum
chaos} point of view: one wants to understand how the classical
dynamics influences the quantum one, both regarding time dependent and
time independent (that is, spectral) quantities. Equivalently, one
searches for traces of classical chaos in the quantum
mechanical system.
\begin{figure}[hp]
\begin{center}
 \includegraphics[width=0.45\textwidth]{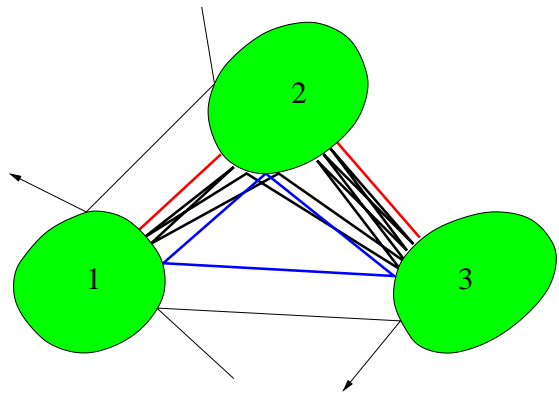}
\caption{Configuration of $3$ convex obstacles in the plane satisfying
  the no-eclipse condition, leading to
a fractal hyperbolic trapped set. The numbering of the obstacles
leads to the associated symbolic dynamics.\label{f:obstacles}}
\end{center}
\end{figure}
In this introduction, I will focus on a
simple system we will be dealing with: the scattering by three or
more balls $(B_j)_{j=1,\ldots,J}$
(more generally, strictly convex bodies with smooth boundaries) in
$\IR^d$, satisfying a no-eclipse condition
\cite{Ikawa88}\footnote{Namely, the convex hull of any pair of
  obstacles $B_i, B_j$ does not intersect any third obstacle.} (see Fig.~\ref{f:obstacles}). The
nature of the classical dynamics will be explained in \S\ref{s:chaotic}.

At the quantum level, one wants to understand the wave
propagation in this geometry, that is solve the (scalar) wave equation
\be\label{e:wave-t}
\partial^2_t u(x,t) - \Delta_\Omega u(x,t) = 0\,,
\end{equation}
with given initial conditions $u(x,0)$, $\partial_t u(x,0)$.
Here $\Delta_\Omega$ is the Laplacian outside the disks
($\Omega=\IR^d\setminus\sqcup_i D_i$), with Dirichlet boundary
conditions. Through a Fourier transform in time, we get the Helmholtz
equation
\be\label{e:Helmholtz}
\Delta_\Omega u (x) + k^2 u(x)=0\,,
\ee
which describes stationary waves of energy $k^2$ ($k$ is the
wavevector, that is the inverse of the wavelength).

If the particle propagating is a quantum one, its evolution rather
satisfies the Schr\"odinger equation 
\be
\label{e:Schrod} i\hbar\partial_t u(x,t)=-\frac{\hbar^2\Delta_{\Omega}}{2}u(x,t)\,,
\ee
where $\hbar$ is Planck's constant. We will see in \S\ref{s:semiclass}
that both equations (\ref{e:wave-t},\ref{e:Schrod}) can be analyzed along the same lines in the high
frequency/semiclassical limits.

This scattering system is physically relevant, and has been studied in
theoretical physics (see for instance the review paper by Wirzba
\cite{wirzba99} for the 2-dimensional scattering by $J$ disks,
and references therein) and mathematics literature \cite{Ikawa88}. It
has also been implemented in various
experimental realizations, most recently on microwave tables by the
Marburg group \cite{Weich10}.

\subsection{Scattering vs. metastable states}\label{s:resonances}
For a given wavevector $k$, Eq.~(\ref{e:Helmholtz}) admits an infinite 
dimensional space of solutions $u(x)$, called \emph{scattering states},
which can be parametrized by decomposing $u(x)$, away from the
obstacles (say, outside a ball $B(0,R_0)$), into a basis of incoming
and outgoing waves:
\be\label{e:in-out}
u(x)= u_{in}(x) + u_{out}(x)\,.
\ee
For instance, in dimension $d=2$ the ingoing waves can be expanded
in angular momentum eigenstates: using polar coordinates $x=(r,\theta)$, 
\be\label{e:u_in}
u_{in}(x)=\sum_{n\in\IZ} a^{in}_{n}e^{i n\theta}\,H^{in}_{n}(kr)\,,
\ee
where $H_{n}^{in}$ are the incoming Hankel functions, and similarly for $u_{out}(x)$.
Any such solution $u(x)$ is called a scattering state. It is
not $L^2$-normalizable, reflecting the fact that the spectrum of
$-\Delta_\Omega$ is absolutely continuous on $\IR_+$, without any embedded
eigenvalue. We will briefly address the phase space structure of these scattering states in \S\ref{s:scatt}.

Beyond the a.c. spectrum, this system admits a discrete set of quantum
{\it resonances}, or complex generalized eigenvalues. They can be
obtained as follows. The
resolvent $(\Delta+k^2)^{-1}$ is a bounded operator on $L^2(\Omega)$
for $\Im k>0$; its norm diverges when $\Im k\to 0$, reflecting the
presence of the continuous spectrum. However, if we cut it off by a
compactly supported (or fast decaying) function $\chi(x)$, the cutoff
resolvent $\chi(\Delta+k^2)^{-1}\chi$ can be meromorphically continued
from the upper half plane $\{\Im k>0\}$ to the lower half
plane $\{\Im k <0\}$, where it generally
admits a discrete set of poles $\{k_j\}$ of finite multiplicities\footnote{In even dimension the continuation
has a logarithmic singularity at $k=0$, often represented by a cut
along the negative imaginary axis.}.
These poles are called the resonances of $-\Delta_\Omega$. Each pole $k_j$
(assuming it is simple) is associated with a generalized eigenfunction $u_j(x)$,
which satisfies the equation
\be\label{e:resonant}
(\Delta_\Omega + k_j^2)u_j(x)=0\,,\quad \text{with Dirichlet boundary
  conditions on $\partial\Omega$}\,,
\ee
and is {\it purely outgoing} (meaning that its decomposition \eqref{e:in-out}
only contains outgoing components). This function grows exponentially
when $|x|\to\infty$, an ``unphysical'' behaviour, so it is
only meaningful inside a compact set (the {\it interaction region}
formed by the ball $B(0,R_0)$). 
The time dependent function 
\be\label{e:metastable}
\tilde{u}_j(x,t) \defeq u_j(x)e^{-ik_jt},\ \ t\geq 0\,,
\ee
satisfies Eq.~\eqref{e:wave-t}. The time decay in \eqref{e:metastable}
explains why $u_j(x)$ is called a {\it metastable state}, with {\it lifetime}
$$
\tau_j = \frac{1}{2|\Im k_j|}\,.
$$
One can expand the time dependent wave $u(x,t)$ when $t\to\infty$ into a
sum over (at least some of) the metastable states
\eqref{e:metastable}. Such
an expansion is less straightforward than in the case of a closed
system (it isn't based on an $L^2$ orthogonal decomposition), but
often gives a good description of the wave $u(x,t)$ for long times \cite{TanZw00,BuZw01,GuiNaud09}. 

Another application of the study of resonances: the presence of a
resonance free strip below the real axis
(together with estimates of the resolvent in the strip) can be used to
quantitatively estimate the
dispersion and local energy decay for the
wave $u(x,t)$, either in the case of the wave equation \eqref{e:wave-t}
or that of the (nonsemiclassical) Schr\"odinger
equation $i\partial_t u=-\Delta_{\Omega} u$ \cite{Chris08,Chris09,BGH10}. 

\subsection{Semiclassical distribution of resonances\label{s:questions}}
We will not investigate these time dependent aspects any further, but
will concentrate on the
spectral one, namely the distribution of the resonances and the
associated metastable states. The first mathematical works on the
subject consisted in counting resonances in large disks $D(0,k)$, $k\to\infty$. 
Melrose \cite{Melrose84} obtained the general Weyl type upper
bound $\cO(k^d)$ for compact obstacles in odd dimension; this bound was generalized to obstacles in even
dimension \cite{Vodev94} as well as to scattering by a potential \cite{Zw89}.

\begin{figure}
\begin{center}
\includegraphics[width=0.7\textwidth]{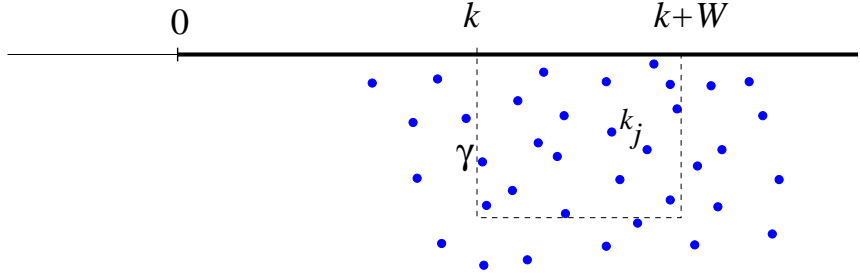}
\caption{Absolutely continuous spectrum of $\Delta_{\Omega}$, together
with the resonances below the real axis, near some value $k\gg 1$.\label{f:reson}}
\end{center}
\end{figure}

In the following we will focus on
the  {\em long living} resonances, that is those $k_j$ sitting within a fixed distance from the real axis
(equivalently, the resonances with lifetimes $\tau_j$ uniformly
bounded from below). These resonances are the most relevant ones
for the long time behaviour of the waves. 
We will consider the high frequency limit $\Re k\gg 1$, which is equivalent with the
semiclassical limit in quantum mechanics (see \S\ref{s:semiclass}), in order to
establish a connection with the classical dynamics (see Fig.\ref{f:reson}). 

\medskip

\noindent{\bf Questions:}\nopagebreak
\begin{enumerate}
\item For given width $W>0$ and depth $\gamma>0$, what is the
  asymptotic number of resonances in the rectangle
$[k,k+W]-i[0,\gamma]$ when $k\to\infty$?
\item In particular, is there some $\gamma>0$ such that this rectangle
  is empty of resonances for $k$ large enough? (such a $\gamma$ is
  called a {\it resonance gap}).
\item Given an infinite sequence of long living resonances $(k_{j_\ell})$, what is the spatial, or phase
  space structure of the associated metastable states when $\Re k_{j_{\ell}}\to\infty$?
\end{enumerate}
In this high frequency limit, these spectral
questions will be connected with long time properties of the classical
dynamics of the system. This dynamics consists in following straight rays at unit speed
outside the obstacles, and reflecting specularly on the obstacles. In
mathematical notations, this dynamics is a {\it flow} $\Phi^t$ defined on
the {\it phase space} formed by the {\it unit cotangent bundle}
$$
S^*\Omega=\{(x,\xi),\,x\in\Omega,\,\xi\in\IR^d,\,|\xi|=1\}\,,$$ 
where the speed $\xi$ is equal to the momentum. 
For each time $t\in\IR$, the flow $\Phi^t$ maps any initial phase
space point $(x,\xi)$ in  to
its position $\Phi^t(x,\xi)$ at time $t$.

With our conditions on the obstacles,
this dynamics is \emph{chaotic} in the following sense: the set of trapped trajectories
\be\label{e:trapped}
\trap = \{\rho\in S^*\Omega,\ \ \Phi^t(\rho)\ \text{uniformly bounded
when $t\to\pm\infty$}\}
\ee
is a fractal flow-invariant set, and the flow on it is uniformly hyperbolic
(equivalently, one says that $\trap$ is a hyperbolic set for
$\Phi^t$, see \S\ref{s:chaotic} for details). For future use we also
define the outgoing ($K^+$) and incoming ($K^-$) tails of the trapped
set,
\be\label{e:tails}
K^{\pm} = \{\rho\in S^*\Omega,\ \ \Phi^t(\rho)\ \text{uniformly bounded when } t\to \mp\infty \}\,,
\ee
with the obvious property $K=K^-\cap K^+$.

The question (2) above has
been addressed around the same time by Ikawa \cite{Ikawa88} and
Gaspard and Rice \cite{GaRi} (see also \cite{Burq93}). 
Both these works establish the presence of a gap,
provided the trapped set $\trap$ is sufficiently ``filamentary''.
The precise criterion depends on a certain dynamical quantity associated
with the flow, a {\it topological pressure} defined in terms of the
{\it unstable Jacobian} $\varphi^+(\rho)$ which measures the instability of
the trajectories (these quantities will be defined in
\S\ref{s:chaotic}, more precisely Eqs.~\eqref{e:pressure} and \eqref{e:varphi+}).
\begin{thm}\label{t:gap-obstacles}\cite{Ikawa88}
Consider the obstacle scattering problem in $\IR^d$, with strictly
convex obstacles satisfying the no-eclipse condition. 
If the topological pressure for the flow on the trapped set $\trap\subset S^*\Omega$
satisfies
$$
\cP=\cP(-\varphi^+/2,\Phi^t\restriction_{\trap})<0\,,
$$
then for any small $\eps>0$ there exists $k_\eps>0$ such that
$\Delta_{\Omega}$ has no resonance  in the strip
\be
[k_\eps,\infty) - i[0,|\cP|-\eps]\,.
\ee
\end{thm}
In dimension $d=2$, the sign of the above topological pressure is
determined by purely geometric data, namely the {\em Hausdorff dimension} of the trapped set:
\be\label{e:pressure-dim}
\cP(-\varphi^+/2,\Phi^t\restriction_\trap) < 0 \Longleftrightarrow \dim_H(\trap)<2\,,
\ee
which gives a precise notion of "filamentary" or "thin" trapped set
(notice that $K$ is embedded in the 3-dimensional phase space $S^*\Omega$).

In \S\ref{s:proof} we will sketch the proof
of the above theorem (in a more general context), using the tool of
quantum monodromy operators. The intuitive idea is the following: wavepackets
propagating along $K$ disperse exponentially fast due to the hyperbolicity of the
trajectories; on the other hand, the wavepackets propagating on nearby
trajectories could also interfere constructively in order to recombine
themselves along $K$. The pressure criterion ensures that the
dispersion is stronger than the possible constructive interference,
leading to a global decay of the wave near $K$.

The pressure $\cP(-\varphi^+/2)$ will appear
several times in the text. 
Its value somehow determines a dichotomy between the
``very open'' scattering systems with ``thin'' trapped sets
($\cP(-\varphi^+/2)<0$), and
the ``weakly open'' ones with ``thick'' trapped sets
($\cP(-\varphi^+/2)\geq 0$). It will be relevant also in the
description of the scattering states in \S\ref{s:scatt}.

\medskip

The question (1) has first been addressed by Sj\"ostrand in the case
of a real analytic Hamiltonian flow with a chaotic trapped set \cite{Sj90}, leading to
the first example of {\it fractal Weyl upper bound}. His result was
generalized and sharpened in \cite{SjZw07}, see Thm~\ref{thm:fractal-h}
below. For the above obstacle scattering, a similar fractal upper
bound had been conjectured in \cite{Sj90}, but proved only recently
\cite{nsz2}.

To state the result, we recall the definition of the upper box (or Minkowski)
dimension of a bounded set $K\subset \IR^{n}$:
$$
\overline{\dim}(K)\defeq\limsup_{\eps\to 0}\Big( n-\frac{\log\Vol(K_\eps)}{\log \eps}\Big)\,,
$$
where $K_\eps$ is the $\eps$-neighbourhood of $K$. The
dimension is said to be {\it pure} if $
\frac{\Vol(K_\eps)}{\eps^{n-\overline{\dim}(K)}}$ is uniformly bounded
as $\eps\to 0$.

\begin{thm}\label{thm:fractal}\cite{nsz2}
Consider the obstacle scattering problem in $\IR^d$, with strictly
convex obstacles satisfying the no-eclipse
condition. 

Let $2\nu+1$ be the upper box dimension of the trapped set
$\trap\subset S^*\Omega$.

Then, the resonances of $\Delta_{\Omega}$ satisfy
the following bound. For any $\gamma>0$ and any $\eps>0$, there exists
$k_{\gamma,\eps},\ C_{\gamma,\eps}>0$ such that
$$
\forall k>k_{\gamma,\eps},\qquad \sharp \{ k_j \in [k,k+1]-i[0,\gamma]\}\leq C_{\gamma,\eps}
k^{\nu+\eps}\,.
$$
If $\trap$ is of {\it pure} dimension $2\nu+1$, one can take $\eps=0$.
\end{thm}
In dimension $d=2$, the trapped set is always of pure dimension, and its box
dimension is equal to its Hausdorff dimension. In that case, the dimension
$\nu$ can be obtained through the topological pressure
of the flow on $K$ (see \S\ref{s:topo-pressure}), namely $\nu$ is
the (unique) real root $s_0$ of the equation
$$
\cP(-s\varphi^+,\Phi^t\rest K)=0\,.
$$
(because $\cP(-s\varphi^+)$ is strictly decreasing with $s$, this
equation directly leads to the equivalence \eqref{e:pressure-dim}).

\medskip

The question (3) has been studied only recently, in the case of a
smooth Hamiltonian flow, or a discrete time dynamics (open
map). The main phenomenon is that, in the high frequency limit, the long living
metastable states are microlocalized
near the outgoing tail $K^+$, and can be described in terms of
semiclassical measures which are invariant through the flow, up to a
global decay (see \S\ref{s:meta}). Although no rigorous result on this
question has been obtained in the case of obstacles, it is very likely
that Thm.~\ref{thm:decay} can be adapted to the obstacles setting.

\subsection{Outline of the paper}

In the next section we
describe the dynamical properties of the classical flows we wish to
consider, namely Hamiltonian flows for which the trapped set is a
compact hyperbolic repeller. We also define the relevant dynamical
quantities associated with the flow, like the unstable Jacobian and
the topological pressure.

In \S\ref{s:semiclass} we 
extend the above two theorems to more
general systems, namely semiclassical Schr\"odinger operators $P(\hbar)$
involving a compactly supported potential, and Laplace-Beltrami
operators on Riemannian manifolds, where
the dynamics is only driven by the geometry. We state the analogues of
Thms.~\ref{t:gap-obstacles} and \ref{thm:fractal} in these settings. 
The case of hyperbolic manifolds of infinite
volume (obtained as quotients of the Poincar\'e half-space
$\mathbb{H}^{d}$ by certain discrete groups) is particularly
interesting: the quantum resonances of the Laplacian can then be directly
connected with the classical dynamics.

In \S\ref{s:massaging} we interpret the quantum resonances 
as the eigenvalues of a (nonselfadjoint) operator obtained by
``deforming'' $P(\hbar)$ into the complex plane. We then (sketchily) explain
how a further deformation, using microlocal weights, allows to prove a fractal Weyl
upper bound for the number of resonances \cite{SjZw07}.

In \S\ref{s:Poincare} we introduce the model of open (chaotic) maps and
their quantizations, which correspond to discrete time dynamics
instead of flows. They have been used as a convenient toy model for
the ``true'' scattering systems, being much more amenable to numerical
studies. We then construct {\it quantum monodromy operators}
associated with a scattering Hamiltonian $P(\hbar)$; they form a
family of open quantum maps associated with the Poincar\'e map
for the classical flow, and can be used to characterize and study the
resonances of $P(\hbar)$.

In \S\ref{e:sharp-FWL?} we formulate a weak and a
strong form of {\it fractal Weyl law}, and discuss their validity
for the various systems introduced
above, mostly guided by numerical data. In \S\ref{s:monodromy-fractal}
we give a heuristic explanation of the
Weyl law for quantum maps, and
provide a proof of its upper bound, eventually leading to
Thm~\ref{thm:fractal} and its analogues. The proof shows how the full quantum
system can be reduced to an effective operator of minimal rank.

In \S\ref{s:gap} we show that the pressure criterion of
Thm~\ref{t:gap-obstacles} applies to all the systems considered
above. We sketch the proof of this gap in the case of
open quantum maps and monodromy operators, leading to the general case of
Schr\"odinger operators. In \S\ref{s:pressure-optimal?} we discuss the sharpness of 
this criterion, using both analytical and numerical results.

In \S\ref{s:metastable} we briefly describe what is known about
the phase space structure of metastable states associated with the
long living resonances, in particular using the tool of semiclassical measures. We
also consider the scattering states. 

Finally, \S\ref{s:concl} presents a brief conclusion, and mentions
possible extensions of the methods to similar nonselfadjoint spectral
problems, like the case of damped waves on a compact manifold of negative curvature.

Most of the above results have appeared elsewhere (or are bound to do
so in a near future). The spectral radius estimate for open
quantum maps, Thm~\ref{t:gap-M}, has not been formulated before, but
it is a rather direct application of \cite{NZ2}. The numerics of
\S\ref{s:radius-baker} had not been published either.

\subsubsection*{Acknowledgements} I have benefitted from many
interesting discussions on this topic, notably with M.Zworski, C.Guillarmou,
F.Naud, M.Novaes, M.Sieber and J.Keating. I
thank C.Guillarmou and F.Naud for communicating to me their recent results on
scattering states. I am also grateful to M.Zworski, M.Novaes
and J.Keating for their permission to use some figures from earlier
publications. I have been partially supported by the
Agence National de la Recherche through the grant
ANR-09-JCJC-0099-01. Finally, I thank both anonymous referees for their
careful reading and constructive comments.

\section{Chaotic dynamics\label{s:chaotic}}

We have already introduced the flow $\Phi^t$ on the phase space $S^*\Omega$ describing
the classical scattering system outside the obstacles: it consists in
the free motion at unit speed outside the obstacles, plus specular reflection at the
boundaries. This flow is generated by the Hamiltonian vector field
$H_p=\frac{\partial p}{\partial\xi}\partial_x - \frac{\partial
  p}{\partial x}\partial_\xi$ associated with the
Hamilton function
\be\label{e:p-obstacle}
p(x,\xi) = \frac{|\xi|^2}{2} + V_{\Omega}(x),\,,\quad\text{with the singular
  potential }\begin{cases}V_{\Omega}(x)=0,&x\in\Omega,\\V_{\Omega}(x)=\infty,&\text{otherwise}\,.\end{cases}
\ee
The unit cotangent bundle $S^*\Omega$ is equivalent with the energy shell
$p^{-1}(1/2)$.

All the flows we will consider are Hamiltonian, in particular they
preserve the natural symplectic form on $T^*\RR^d$. The discrete time
models (open maps) introduced in \S\ref{s:Poincare} will be given by local diffeomorphisms on a
symplectic manifold, which also preserve the symplectic structure. All
these systems are therefore conservative.

We will also make strong dynamical assumptions on these flows (or open
maps), namely
an assumption of \emph{strong chaos}.  The chaotic properties refer to
the long time properties of the
flow restricted to the trapped set $K$ (below we keep the
notations of the obstacle problem, while the concepts apply as well to
more general Hamiltonian flows or to maps). In our setting, chaos is
the mixture of two components, namely hyperbolicity and complexity
(see e.g. the textbooks \cite{KaHa95,BriStu02}).

\subsection{Hyperbolicity}\label{s:hyperb}
\begin{figure}
\begin{center}
\includegraphics[width=.6\textwidth]{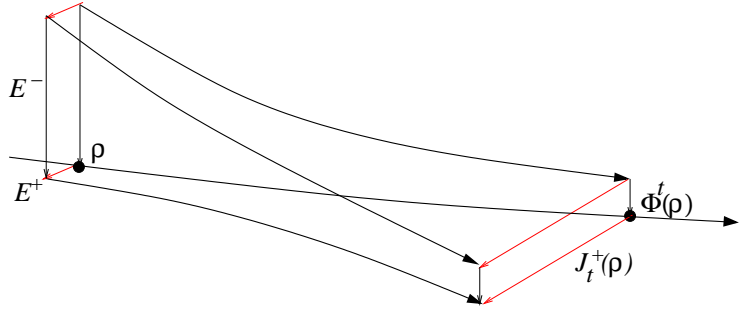}
\caption{Hyperbolicity of the trajectory $\Phi^t(\rho)$: nearby orbits
along the stable (resp. unstable) directions approach $\Phi^t(\rho)$ exponentially
fast in the future (resp. past). The unstable Jacobian $J^+_t(\rho)$
measures this exponential divergence.\label{f:Anosov}}
\end{center}
\end{figure}

Firstly, the trapped set $K$ is assumed to be compact, and the flow
$\Phi^t\rest K$ is assumed to be uniformly
hyperbolic (see Fig.~\ref{f:Anosov}). This means that there is no fixed point ($H_p\neq 0$), and for
any $\rho\in\trap$, the tangent space $T_\rho(S^*\Omega)$ splits into the
flow direction $\RR H_p(\rho)$,
a stable subspace $E^-(\rho)$, and an unstable subspace $E^+(\rho)$: 
$$
T_\rho(S^*\Omega) = \RR H_p(\rho) \oplus E^-(\rho) \oplus E^+(\rho)\,.
$$
The (un)stable subspaces are characterized as follows: there exist $C,\lambda>0$ such that, for any $\rho\in\trap$,
$$
v\in E^{\mp}(\rho)\Longleftrightarrow \forall t>0,\quad \|d\Phi^{\pm t}v\|\leq
C\,e^{-\lambda t}\|v\|\,.
$$
The subspaces $E^{\pm}(\rho)$ depend continuously on
$\rho\in K$, and are uniformly transverse to each other. An important
quantity is the {\it unstable Jacobian} of
the flow,
$$
J^+_t(\rho)\defeq
\det(d\Phi^t\restriction_{E^+(\rho)})\,,\quad t>0\,,
$$
which measures the expansion of the flow along the unstable
manifold\footnote{Although this Jacobian depends on the choice of
  coordinates and metric near $\rho$ and $\Phi^t(\rho)$, its
  asymptotical behaviour for $t\to\infty$ does not.}. This Jacobian grows
exponentially when $t\to\infty$. The infinitesimal
version of this Jacobian reads
\be\label{e:varphi+}
\varphi^+(\rho)\defeq \frac{dJ^+_t}{dt}(\rho)\rest_{t=0}\,,
\ee
and it is possible to choose a metric near $K$ such that $\varphi^+$
is positive on $K$.

The (un)stable subspaces have nonlinear counterparts, namely the
(un)stable manifolds
$$
W^{\mp}(\rho) = \{\rho'\in S^*\Omega,\
{\rm dist}\big(\Phi^t(\rho'),\Phi^t(\rho)\big)\stackrel{t\to\pm\infty}{\longrightarrow} 0\}\,.
$$
The unions of these manifolds makes up the incoming/outgoing tails
$K^{\mp}$ \eqref{e:tails}.

A trapped set $K$ hosting such dynamical properties is called
a \emph{hyperbolic set}, or hyperbolic repeller.

In case of the scattering by $J\geq 2$ convex obstacles, the hyperbolicity
is due to the strict convexity (or positive curvature) of the obstacles, which
defocusses incoming parallel rays at each bounce. On the opposite, in the geometric
scattering models of \S\ref{s:geom}, the defocussing is due to the negative curvature of the manifold.

\subsection{Complexity}\label{s:topo-pressure}

The second ingredient of a chaotic flow is {\it complexity},
in the information theoretic sense. It means that the trapped set $K$
cannot be too simple, e.g it cannot just consist in finitely many periodic trajectories. 
Grossly speaking, complexity means that, if one groups the long
segments of trajectories into ``pencils'' of nearby segments, then the number of
such pencils grows exponentially with the length of the segments. 

Let us make this notion more explicit for our obstacle problem (see Fig.~\ref{f:obstacles}).
To any point $\rho\in K$ away from the obstacle, we can associate a bi-infinite
sequence of symbols 
$$
\bep=\cdots\eps_{-2}\eps_{-1}\cdot\eps_0\eps_1\eps_2\cdots\,,\qquad \eps_i\in\{1,2,\cdots,J\}
$$
indexing the obstacles successively hit by $\Phi^t(\rho)$ in the
future or in the past. This sequence obviously satisfies the condition $\eps_i\neq \eps_{i+1}$ for all $i\in\IZ$.

Conversely, the assumptions we put on the obstacles
imply that, for any sequence satisfying the above condition, 
one can construct a
trajectory with the above properties, and this trajectory is
(essentially) unique. In particular, 
this trajectory is periodic iff the sequence $\bep$ is so. 
This description of the trapped orbits in terms of
sequences of ``symbols'' is called a {\it symbolic dynamics}. It is a
simple way to classify the trajectories of the flow, and estimate its
complexity. For instance, if one decides to group trajectories by
specifying the obstacles they hit from $i=0$ to $i=n$, then the number
of such ``pencils'' is  $J(J-1)^n$, which obviously
grows exponentially with the ``discrete time'' $n$. The number of
periodic orbits also grows exponentially w.r.t. their periods.

This complexity can be made quantitative through the {\it topological
pressures} $\cP(f,\Phi^t\rest K)$ associated with the flow on $K$ and
``observables'' $f\in C(K)$. The pressures provide a
statistical information on the ``pencils'' of long orbit segments.
In the present case of a hyperbolic
repeller, this pressure can be defined in terms of long periodic orbits:
\be\label{e:pressure}
\cP(f,\Phi^t\restriction_\trap)\defeq \lim_{T\to\infty}\frac1T \log
\sum_{p:T-1\leq T_p\leq T} \exp(f(p)),\qquad f(p) = \int_0^{T_p}f(\Phi^t(\rho_p))\,dt\,. 
\ee
The sum runs over all the periodic orbits $p$ of periods
$T_p\in[T-1,T]$, and $\rho_p$ is any point on $p$. 
In \S\ref{t:gap-M} we will give an alternative way to compute the topological
pressure for the open baker's map, in terms of symbolic dynamics.
The pressure can also be defined through a variational formula over
the probability measures $\mu$ on $K$ which are invariant by
the flow:
\be\label{e:variational}
\cP(f,\Phi^t\restriction_\trap)=\sup_{\mu}\Big\{H_{KS}(\mu) + \int f\,d\mu\Big\}\,,
\ee
where $H_{KS}(\mu)$ is the Kolmogorov-Sinai entropy of the measure
$\mu$ (with respect to the flow), a nonnegative number quantifying the
``complexity of a $\mu$-typical trajectory''.

If one takes $f\equiv 0$, the above expression measures the
exponential growth rate of the number of long periodic orbits, which
defines the {\it topological entropy} of the flow:
\be\label{e:top-entropy}
\cP(0,\Phi^t\restriction_\trap) = H_{top}(\Phi^t\restriction_\trap) = \sup_{\mu}{H_{KS}(\mu)}\,.
\ee
For this reason, complexity is often defined by the 
positivity of $H_{top}$.

If $f$ is negative everywhere, the sum in \eqref{e:pressure} shows a competition between exponentially decreasing terms
$e^{f(p)}\sim e^{T\bar{f}}$, and the exponentially increasing number of terms. This is
the case, for instance, if one uses the unstable Jacobian \eqref{e:varphi+} and
takes $f=-s\varphi^+$ for some parameter $s>0$. In that case
$e^{f(p)}=J^+_{T_p}(\rho_p)^{-s}$
measures the instability of the orbit $p$. When $s=1$, the exponential damping exceeds the
exponential proliferation, and the pressure is negative. Actually,
$$
\gamma_{cl}\defeq -\cP(-\varphi^+,\Phi^t\restriction_\trap) > 0
$$
defines the {\it classical decay rate of the flow}, which has the
following physical meaning. Consider an initial smooth
probability measure $\mu_0=g_0(\rho)d\rho$ on $S^*\Omega$, with
the density $g_0$ supported inside the interaction region
$S^*B(0,R_0)$, with $g_0(\rho)>0$ at some point $\rho\in\trap^-$. 
If we push forward this measure through $\Phi^t$, the
mass of the interaction region will asymptotically decay as
\be\label{e:class-decay}
\big[(\Phi^{t})^*\big]\mu_0(S^*B(0,R_0))\sim C\,e^{-t\gamma_{cl}},\quad t\to\infty\,.
\ee
Below we will mostly be interested by the pressure with
observable $f=-\varphi^+/2$. It can be compared with the two quantities
defined above. Indeed, 
using the variational formula \eqref{e:variational} for the pressure,
we easily get
\be\label{e:pressure-convexity}
-\gamma_{cl}/2 \leq \cP(-\varphi^+/2)\leq \frac12 (H_{top}-\gamma_{cl})\,.
\ee
The upper bound embodies the fact that $\cP(-\varphi^+/2)$ is negative if the
dynamics on $\trap$ is ``more unstable than complex''.

In dimension $d=2$ the Hausdorff dimension of $\trap$ can be obtained
in terms of the topological pressure: $\dim_H(K)=2s_0+1$, where 
$s_0\in [0,1]$ is the unique root of the equation
$\cP(-s\varphi^+)=0$. In particular, one gets the equivalence
\eqref{e:pressure-dim}. Hence, hyperbolicity and complexity directly influence the
(fractal) geometry of the trapped set.

\medskip

In \S\ref{s:Poincare} we introduce open maps $\kappa$, which are local
diffeomorphisms defined on some
open subset of a symplectic manifold. The definition of the trapped
set, and of hyperbolicity, are very similar with the case of
flows. Since the Jacobian $J^+_t$ only makes sense for integer times,
we take $\varphi^+=\log J^+_1$. The topological
pressure can be defined as in \eqref{e:pressure}, with $f(p)$
given by a sum over the points in $p$.

\section{Semiclassical formulation and more examples of chaotic flows\label{s:semiclass}}
In the high frequency limit ($k\gg 1$), the Helmholtz equation
\eqref{e:Helmholtz} can be rewritten using an
small positive parameter, which we call $\hbar$ by analogy with
Planck's constant. This parameter scales as
$$
\frac1C\leq \hbar k\leq C\quad\text{(in short, $\hbar\asymp k^{-1}$)},
$$
so the high-frequency limit is equivalent with the
semiclassical limit $\hbar\to 0$. The equation \eqref{e:Helmholtz} now
takes the form of a time-independent Schr\"odinger equation:
$$
-\frac{\hbar^2\Delta_\Omega}{2}u = E(\hbar)u,\quad\text{with
  energy}\quad E(\hbar) = \hbar^2 k^2/2 \in [\frac{1}{2C^2},\frac{C^2}{2}]\,.
$$
Here the operator (quantum Hamiltonian)
\be\label{e:Hamiltonian-obs}
P(\hbar)=-\frac{\hbar^2\Delta_\Omega}{2}
\ee
is the quantization of the classical Hamiltonian \eqref{e:p-obstacle}.

\begin{rem}
The operator $P(\hbar)$ is also the generator of the {\it semiclassical}
Schr\"odinger equation \eqref{e:Schrod}
which describes the scattering of a {\it quantum} (scalar) particle,
$u(x,t)$ being the wavefunction of the particle at time $t$. The
resonant states $u_j(x)$ of \eqref{e:resonant} satisfy the equation 
$P(\hbar)u_j=z_j(\hbar)u_j$, with the
correspondence 
$$
z_j(\hbar)=\hbar^2 k_j^2/2\,.
$$
According to the Schr\"odinger equation \eqref{e:Schrod}, these
states decay with a rate
$|\Im z_j(\hbar)|/\hbar$, which is not the same as the decay rate $|\Im
k_j|$ associated with the wave equation \eqref{e:wave-t}. 
Yet, if we consider resonances in a ``semiclassical box'' $\{\Re (\hbar
k_j) \in [1/C, C],\,\Im k_j\in [-C,0]\}$ for
some fixed  $C>1$, then two decay rates
are comparable: $\Im z_j(\hbar)/\hbar =\Re( \hbar k_j)\,\Im k_j + \cO(\hbar)$.
\end{rem}

\subsection{Potential scattering}
The introduction of $\hbar$ in the
obstacle problem is merely a convenient
bookkeeping parameter in the high frequency
limit. 
More importantly, it allows to extend the our study to
more general scattering Hamiltonian flows, typically by replacing the
obstacle potential $V_{\Omega}$ by a smooth potential $V\in
C^\infty_c(\IR^d)$, leading to the classical Hamiltonian
\be\label{e:classical-Hamil}
p(x,\xi)=\frac{|\xi|^2}{2}+V(x)\,,\quad (x,\xi)\in T^*\IR^d\,,
\ee
which generates a smooth flow $\Phi^t$ on the phase space $T^*\IR^d$.

The $\hbar$-quantization of this Hamiltonian (see Appendix \ref{s:appen}) is the Schr\"odinger operator
\be\label{e:Hamiltonian-V}
P(\hbar)=-\frac{\hbar^2\Delta}{2} + V(x),\quad \,,
\ee
where $\Delta$ is the Laplacian on $\IR^d$. We say that $p(x,\xi)$
is the {\it semiclassical symbol} of the operator $P(\hbar)$ (see the Appendix for a
short reminder on $\hbar$-quantization). If $\supp V$ is contained in a
ball $B(0,R_0)$, we will call this ball the interaction region. 

As opposed to the obstacle case, the flow on the energy shell
$p^{-1}(E)$, $E>0$, is not obtained through a
simple rescaling of the flow at energy $1/2$: both dynamics can be
drastically different. Similarly, at the quantum level, $P(\hbar)$
depends on $\hbar$ in a nontrivial way.
\begin{figure}
\begin{center}
\includegraphics[width=.9\textwidth]{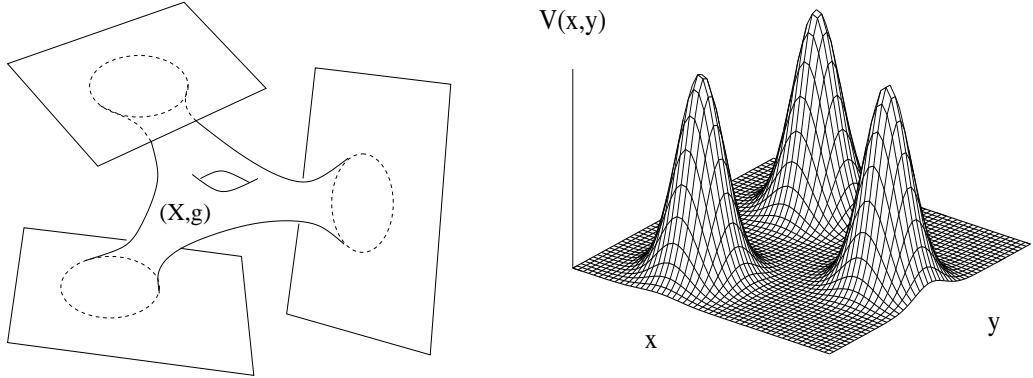}
\caption{\label{f:potential} Left: a Riemannian surface with 3
  Euclidean ends. Right: a potential $ V \in C^\infty_c ( \RR^2 ) $,
with Hamiltonian flow hyperbolic on the trapped set $K_E$ in a range of energies.
}
\end{center}
\end{figure}
It is easy to produce a smooth potential $V(x)$ such that the flow on
the energy shell $p^{-1}(E)$ is chaotic in some range $[E_1,E_2]$,
in the sense that for any energy $E\in [E_1,E_2]$ the trapped set
$$
\trap_E = \{ \rho\in p^{-1}(E),\ \Phi^t(\rho)\ \text{is uniformly bounded for
} t\in\IR\}\,,
$$ 
is a hyperbolic repeller. Following Sj\"ostrand \cite[Appendix
c]{Sj90}, one can for instance ``smoothen'' the hard
body potential $V_{\Omega}$ associated with the above obstacle problem,
and obtain a potential with $J$ ``steep bumps'' (see
Fig.~\ref{f:potential}), such that the flow is chaotic in some
intermediate energy range.

The Schr\"odinger operator \eqref{e:Hamiltonian-V} admits a continuous
spectrum on $\RR_+$, but like in the obstacle problem, its truncated
resolvent
$\chi\big(P(\hbar)-z\big)^{-1}\chi$ can be meromorphically continued from the upper
to the lower half-plane. The poles $\{ z_j(\hbar)\}$ of this continued resolvent 
form a discrete set, which are the resonances of $P(\hbar)$. In
general these resonances depend on $\hbar$ in a nontrivial way. 

We specifically consider the vicinity of a
positive energy $E$ for which the trapped set $\trap_E$ is a
hyperbolic set, and ask the same questions as in \S\ref{s:questions}.
The following results are direct analogues of
Thm~\ref{t:gap-obstacles} and Thm~\ref{thm:fractal} in this semiclassical setting.
\begin{thm}\label{t:gap-semiclass}\cite{NZ2}
Consider the semiclassical Hamiltonian $P(\hbar)$ of \eqref{e:Hamiltonian-V}, such that 
for some energy $E>0$ the flow generated by the
Hamiltonian \eqref{e:classical-Hamil} has a hyperbolic trapped set
$\trap_E$. 

If the topological pressure
$$
\cP=\cP(-\frac12\varphi^+,\Phi^t\rest_{\trap_E}) \quad \text{is negative,} 
$$ 
then for any $\delta,\eps>0$ small enough, there exists
$\hbar_{\delta,\eps}>0$ such that, for any $\hbar\in(0,\hbar_{\delta,\eps}]$ the operator $P(\hbar)$ does not have resonances
in the strip $[E-\delta,E+\delta]-i[0, |\cP|-\eps]$.
\end{thm}
\begin{thm}\label{thm:fractal-h}\cite{Sj90,SjZw07}
Let $P(\hbar)$ be a semiclassical Hamilton operator as in
Thm~\ref{t:gap-semiclass}, and let $2\nu_E+1$ be the upper box dimension of
$\trap_E$.

Then, for any $C>0$, the number of resonances of $P(\hbar)$ in the
disk $D(E,C\hbar)$, counted with multiplicities, is bounded as
follows. For any $\eps>0$, there exists $C_{C,\eps},\
\hbar_{C,\eps}>0$ such that
$$
 \forall\hbar<\hbar_{C,\eps},\qquad
\sharp \big\{ \Res P(\hbar)\cap D(E,C\hbar) \big\} \leq C_{C,\eps}\,
\hbar^{-\nu_E-\eps}\,.
$$
If $\trap_E$ is of pure dimension, one can take $\eps=0$ in the
above estimate.
\end{thm}
One can generalize the above scattering problems on $\RR^d$ by considering
a Schr\"odinger operator $P(\hbar)$ of the form \eqref{e:Hamiltonian-V} 
on an unbounded Riemannian manifold $(X,g)$ with a  ``nice enough'' geometry near
infinity. This geometry should allow to meromorphically extend the
truncated resolvent in some strip. For instance, Thm.~\ref{t:gap-semiclass} applies if
$X$ is a union of Euclidean infinities outside a compact part
\cite{NZ2}. It has been
extended by Datchev \cite{Dat09} and Datchev-Vasy \cite{DatVas10} to more complicated geometries near infinity, in
particular asymptotically Euclidean or asymptotically
hyperbolic manifolds. Their strategy is to ``glue together'' the
resolvent estimates of two model manifolds, one with the true,
trapping structure in the
interaction region but a simple (say, Euclidean) structure
near infinity, and the other one with the true
infinity but a simple (nontrapping) interaction region. In parallel,
Vasy \cite{Vasy10} recently developed a new method to analyze the
resolvent at high frequency, in a variety of asymptotically hyperbolic
geometries.

\subsection{Geometric scattering\label{s:geom}}

On a Riemannian manifold $(X,g)$, the classical scattering in absence of potential
(one then speaks of \emph{geometric scattering}) can already be
complicated, e.g. chaotic. This
is the case, for instance, when the trapped set $K$ lies in
a region of \emph{negative sectional curvature}  (see Fig.~\ref{f:potential},
left). The operator quantizing the geodesic flow is
the (semiclassical) Laplace-Beltrami operator on $X$, $P(\hbar)=-\hbar^2\Delta_X/2$.

One appealing class of examples consists in manifolds
$(X,g)$ obtained by quotienting the Poincar\'e half-space $\IH^{n+1}$
(which has uniform curvature $-1$)
by certain discrete subgroups $\Gamma$ of the group of isometries on
$\IH^{n+1}$.
Such a manifold $X=\Gamma\backslash \IH^{n+1}$ inherits the
uniform hyperbolic geometry of $\IH^{n+1}$, so that all
trajectories are hyperbolic. For a certain type of subgroups $\Gamma$
(called convex co-compact), the manifold $X$
has infinite volume and the trapped set is compact; this trapped set
is then 
automatically a hyperbolic set. This definition of $X$ through group
theory leads to remarkable properties of the spectrum of $\Delta_X$,
which we now summarize (a recent review of the theory in dimension $2$ can be
found in Borthwick's book \cite{Borthwick07}).

The absolutely
continuous spectrum of $-\Delta_X$ consists in
the half-line $[n^2/4,\infty)$, leaving the possibility of 
finitely many eigenvalues in the
interval $(0,1/4)$.
It is common to write the energy variable as
$$
k^2 = s(n-s)\,,\quad \text{so that the a.c.
  spectrum corresponds to } s\in n/2+i\IR\,.
$$
This parametrization has the following advantage. The
resonances of $\Delta_X$, parametrized
by a discrete set $\{s_j\}\subset \{\Re s < n/2\}$, exactly
correspond to the nontrival zeros of the Selberg zeta function \cite{PattPer01}
\be\label{e:Selberg}
Z_X(s) = \exp \Big(  -\sum_p \sum_{m\geq 1} \frac1{m}
\det(1-P_p)^{-1/2} e^{-s m \ell(p)}\Big)\,.
\ee
Here $p$ are the primitive closed geodesics of the flow, and $P_p$ is
the linearized Poincar\'e return map around $p$. This
shows that the (quantum)
resonances are solely determined by the classical
dynamics on $K$. 

One can show that the rightmost zero of $Z_X(s)$
is located at $s_0=\delta$, where $\delta>0$ is the Hausdorff dimension
of the limit set\footnote{Fix one point $x_0\in \IH^{n+1}$. Then the
  limit set $\Lambda(\Gamma)\defeq \overline{\{\gamma\cdot x_0,\
    \gamma\in\Gamma\}}\cap \partial\IH^{n+1}$  actually only depends
  on the subgroup $\Gamma$.} 
$\Lambda(\Gamma)$ \cite{Patt76,Sulliv79}, while all other zeros satisfy 
\be\label{e:X-gap}
\forall j\neq 0,\qquad \Re s_j < \delta\,. 
\ee
If $\delta\in [n/2,n]$ (``thick'' trapped set), $\delta$ corresponds
to the eigenvalue of the ground state
of the Laplacian. There may be finitely many other eigenvalues
($s_j\in [n/2,\delta)$), while the resonances will be located in the
half-space $\{\Re s<n/2\}$.

If $\delta \in (0, n/2)$ (``thin'' trapped set), all the zeros of $Z_X(s)$
correspond to resonances of $\Delta_X$, and since the a.c. spectrum
corresponds to $\Re s=1/2$,  the bound \eqref{e:X-gap}
shows the presence of a resonance gap. 
The topological pressure of $\Phi^t\restriction_{K}$ is given by
$$
\cP(-1/2\varphi^+,\Phi^t\restriction_{\trap})=\delta-n/2\,,
$$
so the bound \eqref{e:X-gap} is a (more precise) analogue of the
semiclassical gaps in Thms~\ref{t:gap-obstacles} and \ref{t:gap-semiclass}.

In this geometric setting one can also obtain fractal upper bounds for the
number of long-living resonances:
\begin{thm}\label{thm:fractal-Gamma}\cite{Zw99,GuiLinZw04}
Let $X=\Gamma\backslash\mathbb{H}^{n+1}$ be a hyperbolic manifold of
infinite volume, with $\Gamma$ a Schottky group\footnote{Schottky
  groups form a certain subclass of convex co-compact groups of
  isometries.}. Then
the resonances of $\Delta_X$ (counted with multiplicities) satisfy 
$$
\forall \gamma>0,\ \exists C_\gamma>0,\ \forall r>1,\quad \sharp \{
s_j \in i[r,r+1] + [n/2-\gamma,n/2] \}\leq C_{\gamma}\,
r^{\delta}\,.
$$
\end{thm}
Notice that $2\delta+1$ is the (Hausdorff or Minkowski) dimension of the
trapped set $K\subset S^*X$.
A slightly weaker result was first obtained in \cite{Zw99} in
dimension 2, using microlocal methods similar with those of
\cite{Sj90} (see \S\ref{s:escape1}). The above result was obtained in \cite{GuiLinZw04} by
analyzing the Selberg zeta function in terms of a {\it classical} expanding map on
$\partial \IH^{n+1}$, and the corresponding transfer operator. This
possibility to rely on purely classical dynamics is specific to the
locally homogeneous spaces $\Gamma\backslash\IH^{n+1}$.

\section{``Massaging''  $P(\hbar)$  into a proof of the fractal
Weyl upper bound\label{s:massaging}}

This section presents two consecutive methods, which were used in
\cite{SjZw07} to prove the fractal Weyl upper bound of Thm~\ref{thm:fractal-h}, that is
in the case of a semiclassical Schr\"odinger operator 
\eqref{e:Hamiltonian-V}. Both methods consist in ``deformations'' of
the original operator $P(\hbar)$, which can be easily analyzed at the level
of the {\it symbols} of the operators, so as to draw consequences on the
spectra of the deformed operators.

The first method, called ``complex scaling'', or rather ``complex
deformation'', provides an alternative definition for the resonances of
$P(\hbar)$, as the eigenvalues of a nonselfadjoint operator.
We remind that resonances were originally
obtained as poles of the meromorphic continuation of the truncated resolvent
$\chi(P(\hbar)-z)^{-1}\chi$. Each
resonance $z_j(\hbar)$ is associated with a metastable state $u_j(\hbar)$, which
is not in $L^2$ but satisfies the differential equation
$P(\hbar) u_j(\hbar)=z_j(\hbar) u_j(\hbar)$.

\subsection{Complex scaling: resonances as spectrum of a
  nonselfadjoint operator\label{s:complex-deform}}

\begin{figure}
\begin{center}
\includegraphics[width=1.\textwidth]{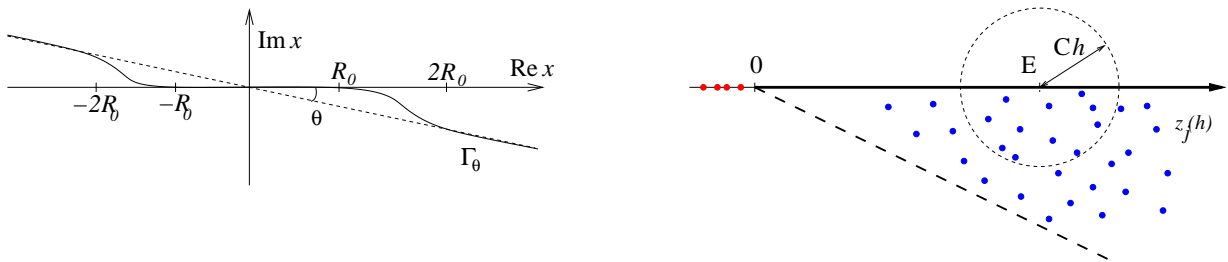}
\caption{Left: deformation of the configuration space $\IR^d$ into a
  contour in $\Gamma_\theta\subset \IC^d$. Right: the spectrum of the
  deformed operator $P_\theta(\hbar)$.\label{f:deformation}}
\end{center}
\end{figure}

The ``complex scaling'' strategy \cite{AgCo,HeSj86} (below we follow
the presentation of \cite{SjZw91}) consists in deforming the configuration
space $\IR^d$ into a complex contour $\Gamma_\theta\subset \IC^d$, $\theta\in[0,\theta_0]$,
of the form
$$
\Gamma_\theta\cap \{|x|\leq R_0\} = \IR^d\cap \{|x|\leq R_0\},\qquad
\Gamma_\theta\cap \{|x|\geq 2R_0\} = \{e^{i\theta}x,\,x\in
\IR^d,\,|x|\geq 2R_0\}\,.
$$
We recall that $B(0,R_0)$ is the interaction region, which contains
the support of the potential.
The differential operator $P(\hbar)$, when analytically extended on
$\Gamma_\theta$, is then equivalent
with an operator $P_\theta(\hbar)$ acting on $\IR^d$. Outside $B(0,2R_0)$
this operator is simply given by $-e^{-2i\theta}\frac{\hbar^2\Delta}{2}$:
this shows that $P_\theta(\hbar)$ is not selfadjoint on $L^2(\IR^d)$, and has
essential spectrum on the half-line $e^{-2i\theta}\IR_+$. More
importantly, its spectrum in the cone $\{-2\theta < 
\arg(z) <  0\}$ is discrete, with eigenvalues $z_j(\hbar)$ equal to the resonances of
the original operator $P(\hbar)$ (see Fig.~\ref{f:deformation}). The associated eigenstates
$u_{j,\theta}(\hbar)$ are equal to the metastable states
$u_j(\hbar)$ inside $B(0,R_0)$, but they are now globally square-integrable.

Through this deformation, the study of resonances
has become a spectral problem for the nonselfadjoint differential operator
$P_\theta(\hbar)$. We can make advantage of
pseudodifferential calculus, that is study the spectrum of
$P_\theta(\hbar)$ by analyzing its semiclassical symbol $p_\theta(x,\xi)$.
For $\theta\ll 1$, this symbol takes the form
\begin{align}
p_\theta(x,\xi) &= p(x,\xi),\quad |x|\leq R_0\\
p_\theta(x,\xi) &= e^{-2i\theta}\frac{|\xi|^2}{2},\quad |x|\geq 2R_0\,.
\end{align}
In particular, for any positive energy $E>0$ and $\delta>0$ small, the
phase space region
$$
V_{\theta}(\delta)\defeq \{(x,\xi)\in T^*\IR^d,\ |p_\theta(x,\xi)-E|\leq\delta\}
$$
is compact (it is contained inside $T^*B(0,2R_0)$). This has
for consequence that, for $\hbar$ small enough, $P_\theta(\hbar)$ has discrete
spectrum in $D(E,C\hbar)$. 
The compactness of $V_\theta(\delta)$ also provides a rough upper
bound on the number of eigenvalues near $E$. Heuristically, the number of
eigenvalues of $P_\theta(\hbar)$ in $D(E,C\hbar)$ is bounded from above by the number of 
quantum states which can be squeezed
in the region $V_{\theta}(C\hbar)$, 
each state occupying a volume $\sim \hbar^d$. This
argument can be made rigorous \cite[Thm~2]{SjZw07}, and produces the bound
\be\label{e:Weyl-coarse}
\#\Spec P_\theta(\hbar)\cap D(E,C\hbar) =\cO\big(\hbar^{-d}\,\Vol
V_\theta(C\hbar) \big) =
\cO(\hbar^{-d+1})\,.
\ee
This estimate does not depend on the nature of the dynamics in the
interaction region. In case the flow on $K_E$ contains stable
orbits surrounded by elliptic islands, one can show that this estimate
is optimal, by explicitly constructing sufficiently many
quasimodes with quasi-energies in $D(E,C\hbar)$ very close to the real
axis, and showing that actual eigenvalues must lie nearby (see
 \cite{TanZw98} and references therein).

\begin{rem}\label{r:absorption}
When solving the
Schr\"odinger equation
$i\hbar\partial_t u = P_\theta(\hbar)u$,
the negative imaginary part of $P_\theta$ acts as an
``absorbing'' term. Indeed, 
a wavepacket $u_0$ microlocalized near a point $(x,\xi)\in
p^{-1}(E)$, $|x|\geq 2R_0$, will be absorbed fast, in the sense that its norm will be
reduced by a factor $\sim e^{t\Im
  p_\theta(x,\xi)/\hbar}=e^{-t\sin(2\theta)E/\hbar}$. Hence, the complex
deformation has the effect to absorb the waves propagating outside the
interaction region.
\end{rem}

The above complex deformation can be used for any type of potential $V(x)$.
In order to refine the counting estimate \eqref{e:Weyl-coarse}, 
one strategy \cite{Sj90,SjZw07} consists in a second deformation of the operator
$P_\theta(\hbar)$, obtained by conjugating
it with an appropriate {\em microlocal weight}, such as to shrink the
region $V_{\theta}(C\hbar)$ to the close vicinity of the trapped
set. Eventhough we will present an alternative (yet related) proof of
Thm~\ref{thm:fractal-h} in \S\ref{s:fractal-proof}, we chose to
sketch this strategy below, which features the power of
pseudodifferential calculus (see the Appendix for a brief introduction).

\subsection{Conjugation by an escape function}\label{s:escape1}
One constructs by hand an {\it escape function} $G\in C_c^\infty(T^*X)$,
which is adapted to the flow $\Phi^t$ in some energy layer
$p^{-1}([E-\delta,E+\delta])$ in the following way. The function $G$
is required to strictly grow along the flow outside an
$\vareps$-neighbourhood $\tilde K_E^\vareps$ of $\tilde K_E\defeq
\cup_{|E'-E|\leq\delta}K_{E'}$ (and for $|x|\leq 2R_0$).
The microlocal weight is then obtained by quantizing this escape
function into the operator $G^w=\Op(G)$, and exponentiating: for some factor $t\gg 1$ one defines the
deformed operator
$$
P_{\theta,tG}(\hbar)\defeq e^{-t G^w}\,P_\theta(\hbar)\,e^{t G^w}\,.
$$
$P_{\theta,tG}$ and $P_{\theta}$ obviously have the same spectrum, but
the pseudodifferential calculus (see the Appendix) shows that the former has a
symbol of the form 
\be\label{e:symbol-p-thetaG}
p_{\theta,tG}=p_{\theta} - i\hbar\, t\, \{p, G\} +\cO((\hbar t)^2)\,,
\ee
where the Poisson bracket $\{p,G\}=H_pG$ is the derivative of $G$ along
the flow generated by $p$. 
From the construction of $G$, this symbol has a
negative imaginary part outside $\tilde K_E^\vareps$,
showing that $P_{\theta,G}(\hbar)$ is absorbing there.
The same volume argument as above then shows that
\be\label{e:Weyl-thin}
\#\Spec P_{\theta,tG}(\hbar)\cap D(E,C\hbar) = \cO\big(\hbar^{-d}\,\Vol V_{\theta,tG}(C\hbar)\big)\,.
\ee 
If $\vareps>0$ is very small, the above bound is sharper than
\eqref{e:Weyl-coarse}, because the set $V_{\theta,tG}(C\hbar)$ has a
much smaller volume than $V_\theta(C\hbar)$. Indeed, the former set is
contained inside  $\tilde K_E^\vareps\cap V_\theta(C\hbar)$, the volume of which scales as\footnote{here we assume
  $\tilde K_E$ is of ``pure'' Minkowski dimension. In the general case one needs to replace $\dim$
  by $\dim +\eps$ for any arbitrary $\eps>0$.}
\be\label{e:fract}
\Vol \Big( \tilde K_E^\vareps \cap \{|p(x,\xi)-E|\leq
C\hbar\} \Big) \asymp \hbar \,\vareps^{2(d-\nu)-2},\qquad \vareps,\hbar \ll 1\,,
\ee
where $2\nu+1$ is the box dimension of $K_E$ inside $p^{-1}(E)$. 
In order to gain a power of $\hbar$ in the right hand side of \eqref{e:Weyl-thin},
we need to take $\vareps\sim\hbar^\alpha$ for some $\alpha>0$, which
implies that the escape function 
$G(x,\xi)$ has to be $\hbar$-dependent. On the
other hand, the pseudodifferential calculus leading to
\eqref{e:symbol-p-thetaG} is valid only if $G$ belongs to a ``good'' symbol class, implying that it cannot fluctuate too
strongly. As explained in Appendix~\ref{s:exotic}, the limiting class corresponds to functions fluctuating on
distances of order
\be\label{e:eps}
\vareps=\vareps(\hbar)\sim\hbar^{1/2}\,.
\ee
Injecting \eqref{e:eps} into \eqref{e:fract} leads to the
fractal Weyl upper bound of Thm~\ref{thm:fractal-h}.

The construction of an optimal escape function $G$
is a bit tricky, it uses the hyperbolicity of the flow on $\tilde
K_E$.
To give a schematic idea, let us consider the simple example of the model Hamiltonian \cite{Christianson07}
$$
p=\xi_1 +  x_2\xi_2\qquad \text{on}\ \ T^*(S^1\times \IR)\,,
$$ 
for which $K_E$ consists in a single hyperbolic periodic orbit $\{\xi_1 = E,\
x_2=\xi_2=0\}$. An ``optimal'' escape function is then
\be\label{e:G_1}
G_1(x,\xi)=\log(\vareps^2 + x_2^2) - \log(\vareps^2 + \xi_2^2) \Longrightarrow
H_p G_1 = \frac{x_2^2}{\vareps^2 + x_2^2} +
\frac{\xi_2^2}{\vareps^2 + \xi_2^2} \,,
\ee
with the scaling \eqref{e:eps}.
Indeed, the gradient $H_pG_1\geq 1$ for $|(x,\xi)|\geq C\vareps$,
while the function remains in a reasonable symbol class.

In the case of a fractal trapped set $K_E$, the escape function $G$ will
locally have a
structure similar with $G_1$ near $K_E$, except that the
coordinates $x_2$, $\xi_2$ are replaced by functions more or less
measuring the distance from the incoming/outgoing tails $K^{\mp}_{E}$.

\section{Open quantum maps and quantum monodromy operators\label{s:Poincare}}

In this section we introduce {\it open quantum maps}, which are toy
models used to study the distribution of quantum resonances, especially in chaotic
situations. These toy models present the advantage to be easy to
implement numerically. 
Besides, the recent introduction of {\it quantum monodromy
  operators} in the context of chaotic scattering establishes a
direct link between the
resonances of a Schr\"odinger operator, and this particular
family of open quantum maps (quantizing
a Poincar\'e return map of the classical flow). 

\subsection{Open quantum maps}\label{s:OQM}

In this section we introduce the {\it open maps}, and their
quantizations, the {\it open quantum maps}.

An open map is a symplectic diffeomorphism
$\kappa:V\mapsto \kappa(V)$, where $V$ and $\kappa(V)$ are bounded
open subsets of a symplectic manifold $\bSigma$, which
locally looks like $T^*\RR^d$. The map $\kappa$ is ``open'' because
we assume that $\kappa(V)\neq V$, so there exist points $\rho\in V$
such that $\kappa(\rho)$ has no further image; we interpret it by
saying that $\kappa(\rho)$ has ``fallen in the hole'', or has
``escaped to infinity'' (this interpretation will become clearer when
we specifically treat Poincar\'e maps). By time inversion, the map
$\kappa^{-1}:\kappa(V)\to V$ is also an open map, and points in
$V\setminus \kappa(V)$ have escaped to infinity ``in the past''.

This escape phenomenon naturally leads to the notions of
incoming/outgoing tails and trapped set: similarly as in
Eqs.~(\ref{e:tails},\ref{e:trapped}), we define
$$
\cK^{\mp}\defeq \{\rho\in V\cup \kappa(V),\ \kappa^{\pm n}(\rho)\in V,\ \forall n > 0 \},\quad \cK=\cK^-\cap \cK^+\,.
$$
We will assume that $\cK$ is compact and at finite distance from the boundary $\partial V$.
As in the case of flows, we will say that the open map $\kappa$ is
{\it chaotic} iff the dynamics generated by $\kappa$ on $\cK$ is
hyperbolic and complex, 
in the sense of \S\ref{s:chaotic}. We will see in \S\ref{s:section} that the Poincar\'e return map of
a chaotic scattering flow is a chaotic open map.
Still, it is not difficult to directly construct chaotic open maps,
e.g. by starting from a chaotic ``closed'' map $\tkappa$ (diffeomorphism) on
$\bSigma$, and restricting it on a subset $V$.

\medskip

What do we call a ``quantization'' of the map $\kappa$? In the case
$\bSigma=T^*\IR^d$, it is a family of operators
$(\cM(\hbar))_{\hbar\to 0}$ on
$L^2(\RR^d)$, with the following asymptotic properties when $\hbar\to
0$.

First, $\cM(\hbar)=\cM(\alpha,\hbar)$ should be an $\hbar$-Fourier Integral Operator (FIO)
associated with $\kappa$, with symbol $\alpha\in C^\infty_c(V)$ (see Appendix~\S\ref{s:FIO} for
more details). For any smooth
observable $a\in C^\infty_c(T^*\IR^d)$, the quantized observable $\Op(a)$ is
transformed as follows when conjugated by $\cM(\hbar)$\footnote{
The notation $A(\hbar)=\cO_{L^2\to L^2}(\hbar^\infty)$ is a shorthand for the fact
that for any $N\geq 0$, $\|A(\hbar)\|_{L^2\to L^2}=\cO(\hbar^N)$ when
$\hbar\to 0$.}:
\be\label{e:Egorov}
\cM(\hbar)^{*}\,\Op(a)\,\cM(\hbar)= \Op(b)+\cO_{L^2\to
  L^2}(\hbar^\infty)\,.
\ee 
The function $b(x,\xi;\hbar)$ is a semiclassical symbol in
$S^0(V)$ supported on $\supp \alpha$, and admitting the
expansion
\be
b= |\alpha|^2\times a\circ\kappa+ \cO(\hbar)\,.
\ee
For $\hbar$ small enough,
$\cM(\alpha,\hbar)$ is uniformly bounded, with
\be\label{e:norm-est}
\|\cM(\alpha,\hbar)\|=\|\alpha\|_{\infty}+\cO(\hbar)\,.
\ee
Equation \eqref{e:Egorov} is a form of Egorov theorem (see \eqref{e:Egorov0}). It implies that
$\cM(\alpha,\hbar)$ transforms a wavepacket $u_0$ microlocalized at a point
$(\rho_0)\in V$ into a wavepacket $\cM(\alpha,\hbar)u_0$ microlocalized at
$\kappa(\rho_0)$, with a norm modified by 
$$
\frac{\| \cM(\alpha,\hbar)u_0 \| }{\|u_0\|} = |\alpha(\rho_0)| + \cO(\hbar)\,.
$$
The coefficient $|\alpha(\rho_0)|$ can be interpreted as an
absorption (resp. gain) factor if $|\alpha(\rho_0)|<1$
(resp.$|\alpha(\rho_0)|>1$). 
The compact support of $\alpha$ shows that the particles outside $\supp\alpha$ are
fully absorbed. 

To call $\cM(\alpha,\hbar)$ an {\it open quantum map}, one furthermore
requires that it is \emph{microlocally unitary} inside $V$. This means
that the above ratio of norms should be $1+\cO(\hbar^\infty)$, for any
state $u_0$ microlocalized inside $V$.
For this to
happen, the symbol $|\alpha(\rho)|$ needs to be a smoothed version of the characteristic function
$\bbbone_V$ on $V$. Typically, one can consider a family of neighbourhoods
$\cK\Subset W_\hbar\subset V$ converging to $V$, say, $W_\hbar = \{\rho\in V,\ dist(\rho,\complement V)\leq
r(\hbar)\}$, e.g. with $r(\hbar)\sim |\log\hbar|^{-1}$, and require that
the symbol $\alpha$ satisfies
\be\label{e:alpha}
\alpha(\rho)= 0 \ \ \text{outside $V$,}\quad |\alpha(\rho)|= 1\ \ \text{inside
$W_\hbar$, and $|\alpha(\rho)|\in [0,1]$ inbetween.}
\ee 
Such $\alpha$ depends on $\hbar$, but in a mild enough way (see the Appendix \S\ref{s:exotic}). To insist
on the regularity of $\alpha$, we will
call such an operator a {\it smooth open quantum map}.

\subsection{Open quantum maps of finite rank\label{s:OQM}}
A priori, the operators $\cM(\alpha,\hbar)$ have infinite rank, even though
they have a finite {\em essential rank}, of the
order of $\hbar^{-d}\Vol(\supp\alpha)$, when $\hbar\to 0$.
This corresponds to the dimension of a subspace 
of states which are not fully absorbed. For practical reasons it can be convenient to
replace $\cM(\hbar)$ by a finite rank operator $M(\hbar)$, by composing $\cM(\hbar)$ with a projector
$\Pi(\hbar)$ microlocally equal to the identity in some neighbourhood of
$\supp\alpha$, and of rank $\sim C\hbar^{-d}$. One then obtains a family of operators
\be\label{e:finite-rk}
M(\hbar) \defeq \cM(\hbar)\,\Pi(\hbar) = \cM(\hbar)+\cO_{L^2\to L^2}(\hbar^\infty)\,.
\ee
Such a projection onto a subspace of finite dimension will be used
in the construction of quantum monodromy operators in \S\ref{s:QMO}.

\medskip

A practical way to construct an open map is to start from a symplectomorphism
$\tkappa$ defined on a compact phase space $\bSigma$, say the
torus $\IT^{2d}=\IR^{2d}/\IZ^{2d}$, and then restrict it to a proper open
subset $V\Subset \IT^{2d}$,
that is take $\kappa = \tilde\kappa\rest_{V}$. There exist recipes to
quantize the ``closed'' map $\tkappa$ into a quantum map \cite{DEGr03}, 
that is a family $(U(\hbar))_{\hbar\to 0}$ of {\em unitary} operators acting
on the family of quantum spaces $(\cH_\hbar)_{\hbar\to 0}$ associated with $\IT^{2d}$, and
enjoying a Egorov property similar with \eqref{e:Egorov}, with
$\alpha\equiv 1$. The spaces $\cH_\hbar$ have dimensions $\sim
(2\pi\hbar)^{-d}$ due to the compactness of $\IT^{2d}$ (with the
constraint $(2\pi\hbar)^{-1}\in\IN$).

To ``open'' this quantum map, one can truncate $U(\hbar)$ by a
projector
quantizing $\bbbone_{V}$, and get the operator
\be\label{e:QM}
M(\hbar) = U(\hbar)\Pi(\hbar)\,.
\ee
The rank of $\Pi(\hbar)$ then scales as
$\hbar^{-d}\Vol(V)$. 
Once again, we speak of a {\it smooth open quantum map} if
$\Pi(\hbar)$ is not a strict projector, but rather a
``quasiprojector'' which is also the
quantization of a ``good'' symbol $\alpha$, like in \eqref{e:alpha}. This
choice allows to avoid diffraction problems near the boundary of $V$.

The above construction was implemented for various chaotic maps on the
2-dimensional torus, chosen such that the open map $\kappa$ admits
a hyperbolic trapped set. In all cases the cutoff $\Pi(\hbar)$
was a sharp projector (in position or momentum).
The first such map was, to my knowledge, the ``kicked rotor'' with absorption \cite{BorGuarShep91}; it was
already aimed at studying the statistics of
quantum lifetimes, defined in terms of the spectrum $\{\lambda_j(\hbar)\}$ of $M(\hbar)$:
\be\label{e:decay-M}
e^{-\tau_j(\hbar)/2} = |\lambda_j(\hbar)|,\qquad j=1,\ldots,\rk(M(\hbar))\,.
\ee
This formula shows that the ``long living'' spectrum of
$M(\hbar)$, say
$\{\lambda_j(\hbar)\}\cap \{|\lambda|\geq r\}$ for some fixed $r>0$, is
seen as a model for the resonances of $P(\hbar)$ in some box
$[E-C\hbar,E+C\hbar]-i[0,\gamma\hbar]$, with the connection $r\equiv e^{-\gamma}$.
Further studies lead to the verification of the fractal
Weyl law \cite{schomerus,NZ1,NZ12} (see
\S\ref{s:maps-fractalWeyl}). 

Such discrete time models have several
advantages. Firstly, the long time dynamics of the classical map $\kappa$ is
sometimes easy to analyze; this is the case for instance for the open
baker's map studied in \cite{NZ1,NZ12}, which we will explicitly describe
in \S\ref{s:baker}. Secondly, the corresponding open quantum maps
are often very explicit matrices, which can be numerically
diagonalized, much more easily so than Schr\"odinger operators like
$P_\theta(\hbar)$. A variant of the quantum baker's map even lends
itself to an analytical treatment (see \S\ref{s:Walsh}).
Third, the
quantum monodromy operators  establish a connection between a family of open
quantum maps and a ``physical'' scattering flow.
(see \S\ref{s:QMO}). To explain the construction of the monodromy
operators we need to recall
the definition of Poincar\'e sections associated with a Hamiltonian flow.

\subsection{Poincar\'e sections}\label{s:section}
\begin{figure}[ht]
\begin{center}
\includegraphics[width=0.32\textwidth]{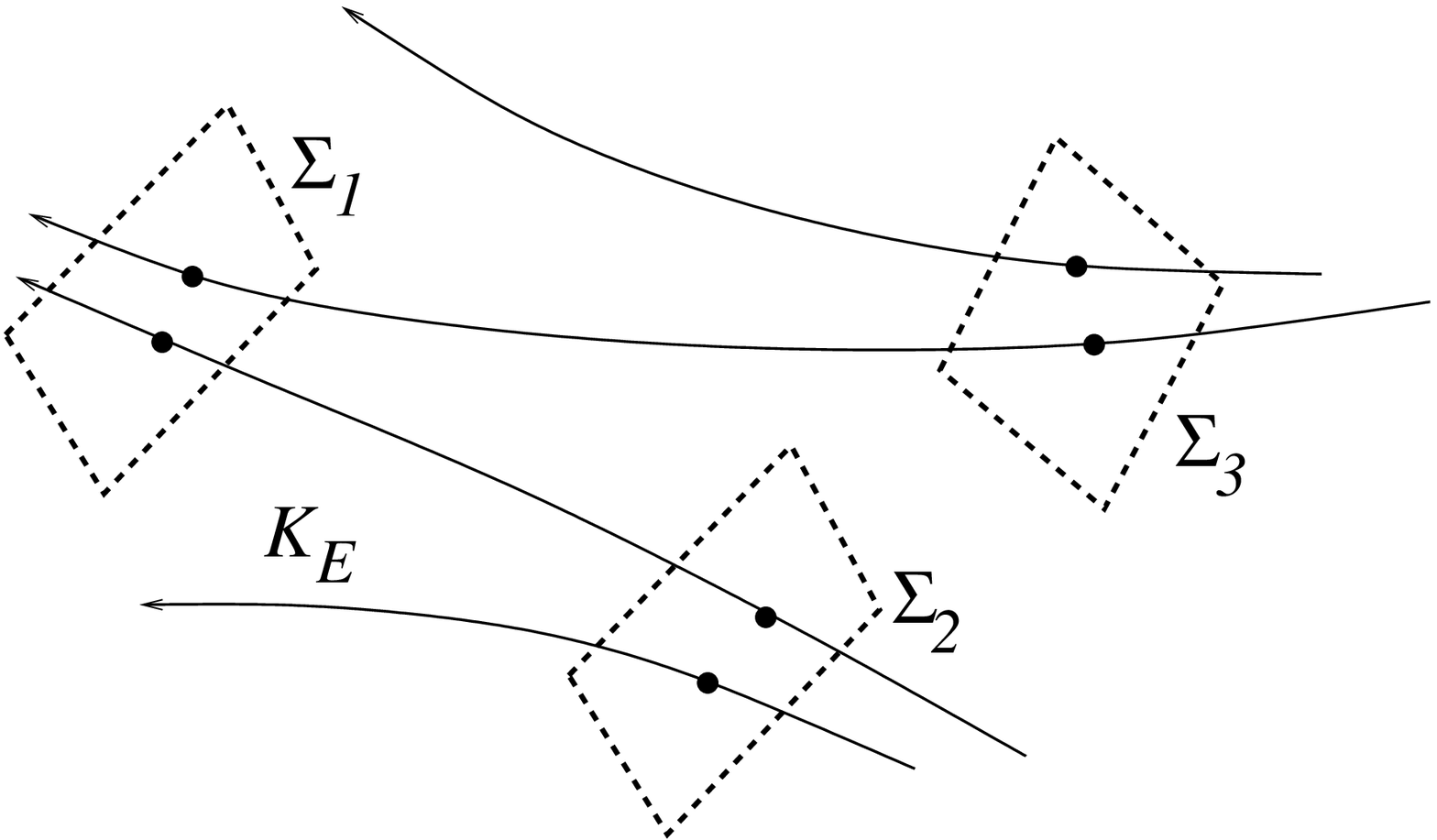}\hspace{1cm}\includegraphics[width=.6\textwidth]{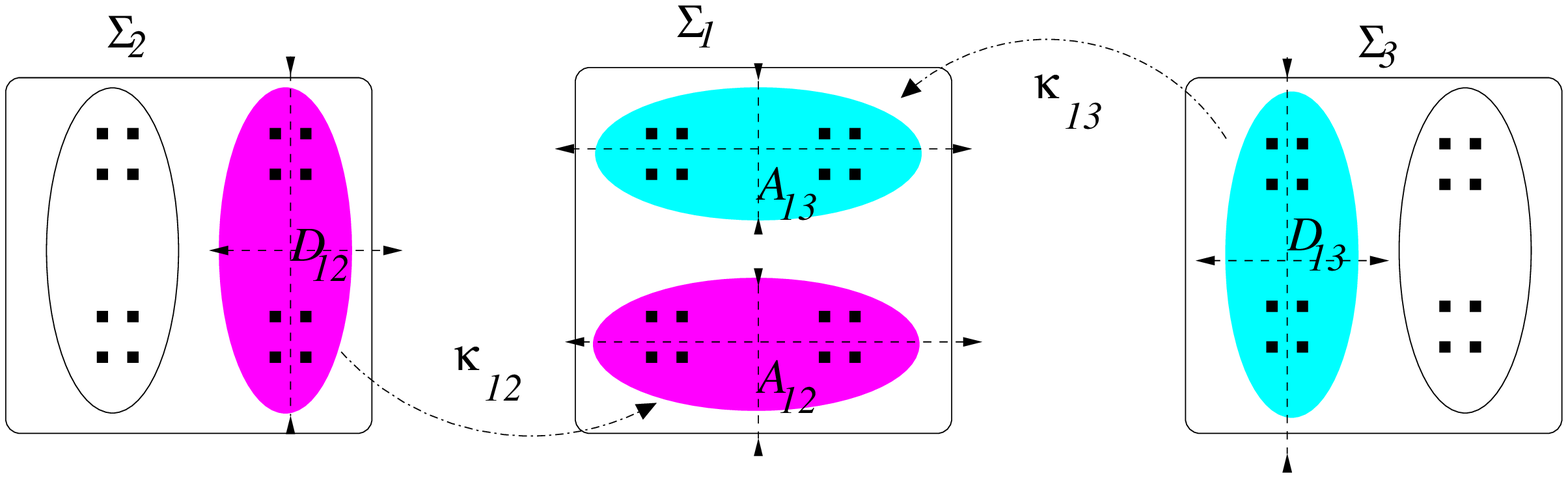}
\caption{Left: schematic representation of a Poincar\'e section $\bSigma$ near the
  trapped set $K_E$. Right: the induced return map $\kappa=(\kappa_{ij})$ on
  $\bSigma$. Vertical/horizontal axes indicate the stable/unstable
  directions, and the trapped set $\cK$ is sketched by the black squares.\label{f:Poincare}}
\end{center}
\end{figure}
We are back to the setting of \S\ref{s:chaotic}, with a Hamiltonian flow $\Phi^t$
on $T^*\IR^d$ (more generally $T^*X$ for some manifold $X$). 
Given $E>0$ a noncritical energy, a Poincar\'e section near the trapped set $K_E$
is a finite union of hypersurfaces $\bSigma=\sqcup_{i=1}^I\Sigma_i$ in $p^{-1}(E)$, uniformly transverse to the
flow, such that for each point $\rho$ sufficiently close to $K_E$ the trajectory $\Phi^t(\rho)$ intersects
$\bSigma$ (in the future and the past) after a uniformly bounded time (Fig.~\ref{f:Poincare}).
This property allows to define 
$$
\text{a return map }\kappa :V\subset \bSigma\to \kappa(V)\subset\bSigma,\qquad \text{and a
  return time }\ \ \tau:V\subset\bSigma\to\IR_+\,,
$$ 
where $V$ is a neighbourhood of the \emph{reduced trapped set}
$$
\cK\defeq K_E\cap \bSigma = \sqcup_{i=1}^I\cK_i\,.
$$ 
Since the flow $\Phi^t$ is symplectic on $T^*X$, the Poincar\'e section $\bSigma$ can be
given a natural symplectic structure, which is preserved by
$\kappa$. Notice the dimensional reduction: $\bSigma$ has dimension $2d-2$.
The flow $\Phi^t$ in the neighbourhood of $K_E$ is fully
described by the pair $(\kappa,\tau)$. In
particular, $\kappa\rest_{\cK}$ is
hyperbolic if $\Phi^t\rest_{K_E}$ is so. Analyzing the ergodic properties of such a hyperbolic
map has proved easier than directly analyzing the flow. Indeed,
the thermodynamic formalism, which allows to construct nontrivial
invariant measures, and analyze their ergodic properties, is based on
such a Poincar\'e reduction \cite{BowRue75}.

Since $\bSigma$ is a union of disjoint hypersurfaces $\Sigma_i$
locally equivalent with $T^*\IR^{2d-2}$, the map $\kappa$ can be seen
as a collection of symplectic maps
$\kappa_{ij}:D_{ij}\subset\Sigma_j\mapsto \Sigma_i$, where
$D_{ij}$ consists of the points in $\Sigma_j$, the trajectories of which next intersect
$\Sigma_i$.
An open quantum map associated with $\kappa$ is then an operator valued matrix
$\cM(\hbar)=(\cM_{ij}(\hbar))_{i,j=1,\ldots,I}$, such that $\cM_{ij}(\hbar)=0$ if
$D_{ij}=\emptyset$, otherwise $\cM_{ij}(\hbar)$ is an open map quantizing
$\kappa_{ij}$. The operator $\cM(\hbar)$ acts on
$L^2(\IR^{d-1})^{I}$, and satisfies a vector-valued Egorov property
similar with \eqref{e:Egorov}, where
$a=(a^i)_{i=1,\ldots,I}$ and $b=(b^i)_{i=1,\ldots,I}$ are observables on $\bSigma$.
One can also consider open quantum maps of finite rank $M(\hbar)$, as in \eqref{e:finite-rk}.

\subsection{Quantum monodromy operators}\label{s:QMO}

It turns out that, under a mild condition on the
trapped set $K_E$, there exists a family of
quantum maps (more precisely, of FIOs) associated with the Poincar\'e map $\kappa$, which allows
to directly recover the resonance spectrum of the quantum Hamiltonian
$P(\hbar)$. 
\begin{thm}\label{thm:QMO}\cite{nsz1,nsz2}
Let $P(\hbar)$ be as in Thm.~\ref{t:gap-semiclass}, and assume
that the trapped set set $K_E$ is totally disconnected transversely to the
flow\footnote{This condition is probably generic within the
family of chaotic scattering flows we are considering. It can be
relaxed a bit, see \cite{nsz1}}.

Alternatively, take $P(\hbar)=-\frac{\hbar^2\Delta}{2}$ the Dirichlet
Laplacian outside $J$ convex obstacles satisfying the no-eclipse
condition, and take $E=1/2$.

Consider a Poincar\'e section $\bSigma=\sqcup_{i=1}^I\Sigma_i$ transverse to
$\Phi^t$ near $K_E$. Then, there exists a family of \emph{quantum
  monodromy operators}
$(M(z,\hbar))_{z\in D(E,C\hbar)}$ on $L^2(\IR^{d-1})^I$, with the following properties:

\noindent(i) $M(E,\hbar)$ is an open quantum map quantizing $\kappa$, of finite
rank $\asymp \hbar^{-d+1}$.\\
(ii) $M(z,\hbar)$ depends holomorphically in $z\in D(E,C\hbar)$, and
\be\label{e:z-dependence}
M(z,\hbar)=M(E,\hbar)\,\Op(e^{-i(z-E)\tau/\hbar}) + \cO(\hbar^{1-\eps})\,,
\ee
where $\tau$ is the return time (smoothly continued outside $V$). \\
(iii) the resonances of $P(\hbar)$ in $D(E,C\hbar)$ are the roots
(with multiplicities) of the equation
\be\label{e:implicit}
\det(1-M(z,\hbar))=0\,.
\ee
\end{thm}
The properties $(i),(ii)$ ensure that for all $z\in D(E,C\hbar)$,
$M(z,\hbar)$ remains an FIO associated with $\kappa$, but for
$z\not\in \IR$ it is no more unitary near $\cK$.

The crucial property $(iii)$ exhibits the
connection between the spectrum of $M(z,\hbar)$ and the resonances of
$P(\hbar)$. It has transformed a linear spectral problem
$(P_\theta(\hbar)-z)u=0$, into a problem $M(z,\hbar)v=v$
of finite rank, depending nonlinearly in $z$.
\begin{figure}[ht]
\begin{center}
\includegraphics[width=0.5\textwidth]{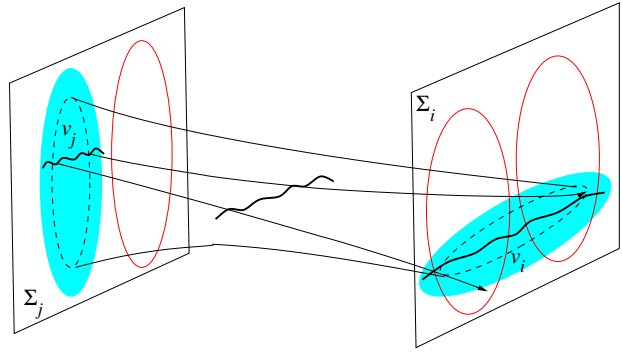}
\caption{Schematic construction of the monodromy operator
  $M_{ij}(z,\hbar)$, by ``following'' a microlocal solution of
  $(P(\hbar)-z)u=0$ from a neighbourhood of $\Sigma_j$ to a
  neighbourhood of $\Sigma_i$.\label{f:monodromy}}
\end{center}
\end{figure}
The construction of the monodromy operators $M(z,\hbar)$ is not
unique, and rather implicit. Roughly speaking, each component
$M_{ij}(z,\hbar)$ is obtained by expressing the microlocal
solutions to the equation $(P_\theta(\hbar)-z)u=0$ near
$\cK_j$ in terms of their ``local transverse data'' $v_j\in
L^2(\IR^{d-1})$, using a choice of coordinates near $\Sigma_i$. The
microlocal solutions $u$ can be continued up to $\Sigma_i$, where they
are analyzed in terms of local transverse data
$v_i$. $M_{ij}(z,\hbar)$ is defined as
the operator mapping $v_j$ to $v_i$: this explains the
denomination of ``monodromy operator'' (see
Fig.~\ref{f:monodromy}). The main technical difficulty
consists in transforming this microlocal characterization into a
globally defined, finite rank operator.

\medskip

A monodromy operator had been constructed, and used to study the
resonance spectrum of a scattering operator $P(\hbar)$, in the case where the trapped set
consists in a single hyperbolic orbit \cite{GeSj87}. In a different
framework, a microlocal
form of monodromy operator associated with an isolated periodic orbit was used
in \cite{SjZw02} to compute the contribution of this orbit to Gutzwiller's
semiclassical trace formula.

In the physics literature, Bogomolny \cite{Bogo92} formally defined a
``quantum transfer operator'' $T(E,\hbar)$ associated with the 
Hamiltonian $P(\hbar)$ of a closed system: this operator quantizes the return map
through a certain spatial hypersurface, and in the semiclassical limit
the eigenvalues of $P(\hbar)$ are (formally) given by the roots of the
equation $\det(1-T(E,\hbar))=0$. This approach was adapted by
Doron and Smilansky to study
the spectrum of closed Euclidean billiards \cite{DorSmil92}, and was
also implemented in a
nonsemiclassical setting by Prosen \cite{Prosen95}.

In the case of the
scattering by $J$ convex obstacles, similar operators were constructed
\cite{Ge88,Ikawa88}. Here the section
$\bSigma$ is ``selected'' by the setting: it
consists in the union of the cotangent bundles of the obstacle
boundaries, $\Sigma_i=T^*\partial\cO_i$. 
To each obstacle $\cO_i$ one associates a
Poisson operator $H_i(k):C^\infty(\partial\cO_i)\mapsto
C^\infty(\IR^d\setminus\cO_i)$, such that
$$\forall v\in C^\infty (\partial\cO_i),\quad 
\begin{cases}(\Delta + k^2)H_i(k)v = 0\ &\text{on
  }\IR^d\setminus\cO_i,\\
H_i(k)v\ \ \text{is outgoing,}&\\
(H_i(k)v)\rest_{\partial\cO_i}=v\,.&
\end{cases}
$$
Then, the scattering problem by the $J$ obstacles can be expressed in
terms of the ``quantum boundary map''
$\cM(k)=(\cM_{ij}(k))_{i,j=1,\ldots,J}$ defined by
$$
\cM_{ij}(k):C^\infty(\partial\cO_j)\mapsto
C^\infty(\partial\cO_i)\,,\qquad
\begin{cases}\cM_{ij}(k)=0,&i=j\\\cM_{ij}(k)\,v=(H_j(k)\,v)\rest_{\partial\cO_i}\,&i\neq j\,.\end{cases}
$$
In the high frequency limit
$k\to\infty$, and away from the ``glancing orbits'', the operator $\cM_{ij}(k)$
has the structure of an open quantum map associated with the boundary map of the
billiard flow, . In \cite{nsz2} we show
how to reduce these boundary operators $\cM(k)$ to finite rank
monodromy operators $M(k)$, which have the properties expressed in the
above theorem (as explained in \S\ref{s:semiclass}, the correspondence with the semiclassical formalism
reads $\hbar\sim |k|^{-1}$, $z=\hbar^2k^2/2=1/2+\cO(\hbar)$).

\section{From  fractal Weyl upper bound to fractal Weyl law ?\label{e:sharp-FWL?}}

In our attempts to address the question (1) in \S\ref{s:questions}, we
have so far only obtained upper bounds for the number of
resonances. Lower bounds are more difficult to derive, due to the fact
that the spectral problem we are dealing with is effectively nonselfadjoint.
Upper bounds are generally obtained by first counting
the {\it singular values} of some operator related with
$P_\theta(\hbar)$, which is a selfadjoint spectral problem; after
controlling the distribution of
singular values one can then apply Weyl's inequalities\footnote{Let
  $(\lambda_i)$ (resp. $(s_i)$) be the eigenvalues (resp. singular
  values) of a compact operator, ordered by decreasing
  moduli. Then, for any $j\geq 1$, $\sum_{i=1}^j |\lambda_i|\leq
  \sum_{i=1}^j s_i$.} to bound (from
above) the number of eigenvalues of $P_\theta(\hbar)$.

The difficulty to obtain a lower bound (that is, ensure that there are
indeed about as many eigenvalues as what is permitted by the upper bound) may be traced to the possible
high sensitivity of the spectrum w.r.t. perturbations.
So far, the only access to lower bound 
is provided by some form of Gutzwiller's (or Selberg's) trace formula.
Using this strategy, lower bounds on the number of resonances
have been obtained in the case of convex co-compact manifolds
$X=\Gamma\backslash\IH^{n+1}$ (we use the notations of \S\ref{s:geom}).
\begin{thm}\cite{GuiZw99,Perry03}
Let $X=\Gamma\backslash\IH^{n+1}$, with $\Gamma$ a convex co-compact and
torsion-free subgroup. Then, for any small $\eps>0$, there exists
$\gamma_\eps>0$ such that\footnote{the notation $f(r)=\Omega(g(r))$ means that
$\frac{f(r)}{g(r)}$ takes arbitrarily large values when $r\to\infty$.}
$$
\sharp \{ s_j \in i[0,r] + [-\gamma_\eps,n/2] \} =
\Omega(r^{1-\eps})\qquad\text{when }r\to\infty\,.
$$
\end{thm}
The proof of this lower bound uses an exact, Selberg-like trace formula, which
connects the resonance spectrum on one side, with a sum over the closed
geodesics on the other side. Applying a well-chosen test function on this trace formula,
one exhibits a singularity on the ``geodesics side'', which implies
(on the ``spectral side'') the presence of
many resonances. 

\begin{figure}[ht]
\begin{center}
\includegraphics[width=0.6\textwidth]{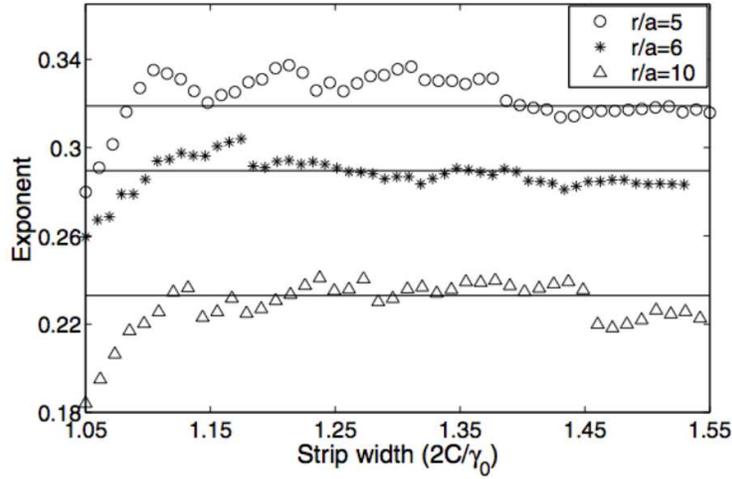}
\caption{Check of the fractal Weyl law for the the scattering by $3$
  disks of radii $a$ located in an equilateral triangle of sidelength
  $r$, for $3$ values of the ratio $r/a$. In each case a fractal exponent
  $\nu(C)$ was extracted from counting the resonances in a long strip
  of depth $C$ for various values of $C$ larger than $2\gamma_{cl}$,
  and compared with the geometrical exponent $\nu$
  (horizontal lines).
Reprinted figure with permission from
\href{http://link.aps.org/abstract/PRL/v91/e154101}{W.~Lu, S.~Sridhar,
  M.~Zworski, Phys. Rev. Lett. {\bf 91}, 154101 (2003)}. Copyright 2003 by the American Physical Society. \label{f:lsz}}
\end{center}
\end{figure}

We notice a gap between this lower bound and the upper bound of
Thm~\ref{thm:fractal-Gamma}, which implies that the left hand side is
bounded above by
$C_\gamma\,r^{1+\delta}$, with $\delta>0$ the dimension of the limit
set. In \cite{GuiZw99} the authors conjecture that the
actual number of resonances in the strip is of the order of the
fractal upper bound. Similar conjectures for various other systems
can be split into two forms.
\begin{defn}Let $P(\hbar)$ be a Schr\"odinger operator as in
  \eqref{e:Hamiltonian-V}, and assume that for some $E>0$ the
  trapped set $K_E$ is a hyperbolic repeller of pure Minkowski
  dimension $1+2\nu$. 
 We define the weak, resp. strong form of {\it fractal Weyl law conjecture} as follows.

\noindent (i) {\bf Weak form.}
For $C,\gamma>0$ large enough (for a very weak form, take $C\sim
\delta \hbar^{-1}$), there exists $C_\gamma>0$, $\hbar_{C,\gamma}>0$ such that
$$
\sharp \{ \Res P(\hbar)\cap [E-C\hbar,E+C\hbar]-i[0,\gamma\hbar]\}
\geq C\,C_\gamma\,\hbar^{-\nu}\,,\quad \forall \hbar<\hbar_{C,\gamma}\,.
$$
(ii) {\bf Strong form.} There exists an increasing function
$F:\IR_+\mapsto \IR_+$, nonidentically vanishing, such that, for any $C,\gamma>0$,
\be\label{e:FWL-strong}
\sharp \{ \Res P(\hbar)\cap [E-C\hbar,E+C\hbar]-i[0,\gamma\hbar]\}
= C\, F(\gamma)\,\hbar^{-\nu}+o(\hbar^{-\nu}),\quad \text{when}\ \hbar\to 0\,.
\ee
\end{defn}
In order to test either form of the conjecture, resonance spectra have been
numerically computed for the three types of systems: the 3-bump
potential \cite{Lin02,LinZw02} or a modified H\'enon-Heiles Hamiltonian
\cite{Ramil+09}, 
the 3-disk scattering on the
plane\cite{LuSrZw03} (see Fig.~\ref{f:lsz}), or a scattering by 4
spheres on the 3-dimensional space \cite{EberMainWunn10}, and several
convex co-compact surfaces \cite{GuiLinZw04}\footnote{In the last
two cases, resonances were obtained by computing the
zeros of the Selberg/Gutzwiller zeta functions: this procedure exactly provides the
resonances in the convex co-compact case, while in the obstacle case the
zeros are believed to be good approximations of
the actual resonances.}. 
In all cases, the
counting was compatible with the fractal Weyl law, although the
convergence to the asymptotic behaviour was difficult to ascertain. 
More recently, attemps have been made to extract the long living
resonances of the 3-disk scattering system from an {\it experimental
  signal} on a microwave (quasi)-2d billiard by the
Marburg group \cite{Marburg08}. Yet, computing high frequency resonances in such an experiment
presents many difficulties: a noisy and discrete
signal, the presence of antennas perturbing the ideal system, the
difficulties to reach sufficiently high frequencies, and
the delicate implementation of the
harmonic inversion method used to extract the ``true'' resonances.

The fractal Weyl conjecture was actually much easier to test
numerically for the toy model of {\em open quantum maps}.

\subsection{Fractal Weyl law for open quantum maps}\label{s:maps-fractalWeyl}

As explained above, in the quantum map framework the distribution of long-living states
is studied by fixing some radius $r>0$, and counting the number of
eigenvalues $\lambda_j(\hbar)$ of $M(\hbar)$ in the annulus $\{r\leq |\lambda|\leq
1\}$. This task is easy to implement numerically for operators
(matrices) $M(\hbar)$ of reasonable dimensions. 
Schomerus and Tworzyd\l o implemented it on the
kicked rotor \cite{schomerus} in a strongly chaotic
r\'egime\footnote{As far as I know, the chaoticity of the kicked rotor has not been
proved rigorously, but seems plausible in view of numerics.}, with a sharp
opening along a vertical strip. A very good agreement with the strong fractal
Weyl law was observed: most eigenvalues
accumulate near $\lambda=0$, while a small fraction of them have
moduli $\geq r$. Their numerics hint at the
existence of a nontrivial profile function $r\in(0,1]\mapsto
F(r)\geq 0$, such that
\be\label{e:FWL-maps}
\forall r>0,\qquad \sharp \big\{ \Spec M(\hbar)\cap\{ |\lambda|\geq r\}\big\}= F(r)\,\hbar^{-\nu}+o(\hbar^{-\nu})\,,
\ee
with $\nu=\dim \cK/2$. 
\begin{figure}
\begin{center}
\includegraphics[width=0.5\textwidth]{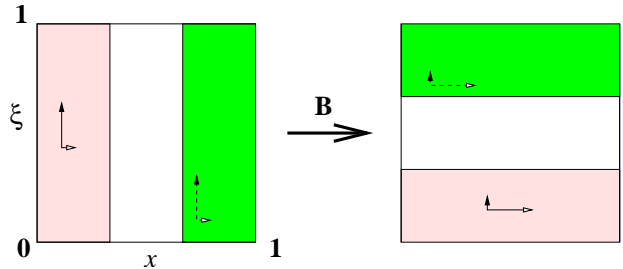}
\caption{Sketch of the symmetric $3$-baker's map, with a
  hole in the central rectangle.
\label{f:baker}}
\end{center}
\end{figure}

\subsubsection{The open baker's map\label{s:baker}}
The spectra of several types of quantum open baker's maps
were analyzed in \cite{NZ1,NZ12,NoRu07}. Let us recall the definition
and basic properties of this family of chaotic maps on $\IT^2$. 

A baker's map $\tkappa$ is defined by splitting $\IT^2$ into $D\geq 2$
``Markov rectangles''
$R_i=\{x_i\leq x<x_{i+1},\,0\leq \xi<1\}$, $i=0,\ldots,D-1$,
$x_0=0,\,x_D=1$, and
mapping the points in $R_i$ as follows:
\be\label{e:baker}
(x,\xi)\in R_i \mapsto \tkappa(x,\xi) = (\frac{x-x_i}{\ell_i},\ell_i
\xi + x_i)\,,\quad\text{with }\ell_i\defeq x_{i+1}-x_i\,.
\ee
Since $\ell_i<1$, the stable/unstable directions are the vertical/horizontal
axes. This map is invertible, but discontinuous along the boundaries $\partial
R_i$. There is an obvious symbolic dynamics: to a point $\rho=(x,\xi)$
one associates the sequence $\cdots\eps_{-1}\cdot\eps_0\eps_1\cdots$,
$\eps_j\in\{0,\ldots,D-1\}$, such that $\tkappa^j(\rho)\in R_{\eps_j}$
for each time $j\in\IZ$. Conversely, to any bi-infinite sequence
corresponds a point $\rho$, and this map is ``almost''
one-to-one\footnote{The defect of injectivity comes from points with
  sequences ending by infinite strings of $0$, on either end. 
For instance, the point $(0,0)$ can be represented by the constant
sequences $\overline{0}$ or $\overline{D-1}$.}.

In order to take advantage of this symbolic dynamics, the opening
$\IT^2\setminus V$ was chosen to consist in the union of $D-n$ of the Markov
rectangles, $0<n<D$. The trapped set $\cK$ is then easy to describe
(see Fig.~\ref{f:baker-trapped}):
it consists in the sequences $\bep$ with all $\eps_j\in
I=\{i_1,\ldots,i_n\}$ the set of ``kept rectangles''. This set is the
cartesian product ${\rm Can}\times {\rm Can}$, where ${\rm Can}$ is a
Cantor set on the unit interval; the Hausdorff or Minkowski dimension $\nu$ of ${\rm Can}$ is
explicitly given by the only real root of the equation
$$
\ell_{i_1}^s + \ell_{i_2}^s +\cdots + \ell_{i_n}^s = 1\,.
$$
For instance, if we keep $n$ rectangles in the symmetric $D$-baker's
map, we get $\nu = \frac{\log n}{\log D}$.

Assume
the $\ell_i$ are rational. For quantum dimensions
$N=(2\pi\hbar)^{-1}$ such that $N\ell_i$ are all integer, the ``closed
map'' $\tkappa$ is
quantized according to the recipes of Balasz-Voros or Saraceno
\cite{BaVo89,Sar90}, namely by the unitary matrix
$$
U(\hbar) = U_N = F_N^{-1}\begin{pmatrix} 
  F_{N\ell_{0}}&&\\& \ddots &\\&&F_{N\ell_{D-1}}\end{pmatrix}\,,
$$
where $F_*$ is the $*$-dimensional discrete Fourier transform. The
quantization of the open map $\kappa$ is simply obtained by projecting
out the $D-n$ blocks $F_{N\ell_j}$ corresponding to the opening.

\begin{figure}
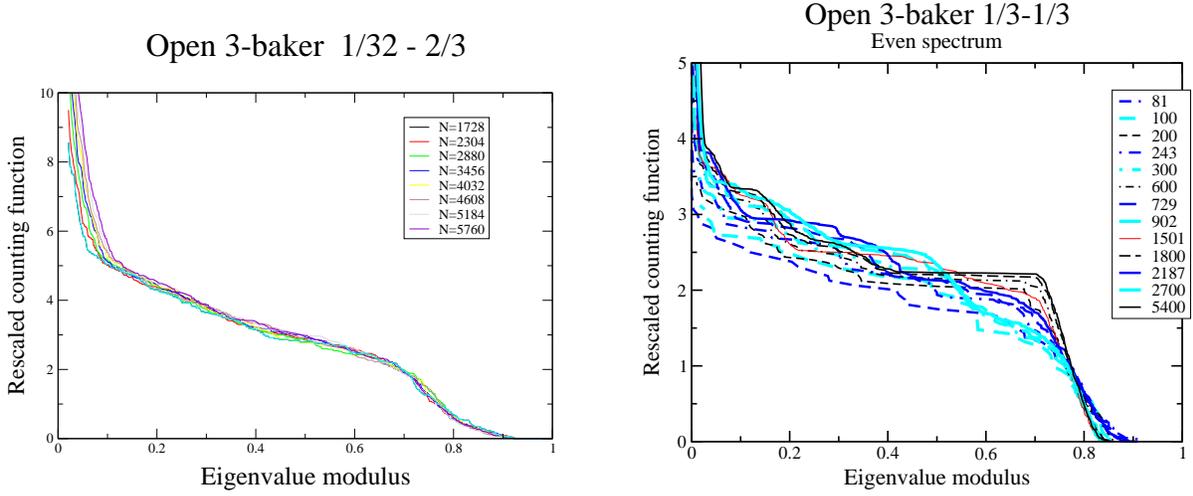

\begin{center}
\includegraphics[width=0.45\textwidth]{spect-asymB-rescaled.eps}\hspace{1cm}
\includegraphics[width=0.45\textwidth]{spect-3e-rescaled-landscape.eps}
\end{center}
\caption{Rescaled counting function for an asymmetric (left) and a
  symmetric (right)
  open 3-baker opened by removing the central rectangle (the fractions
  in the title denote the contraction factors $\ell_0,\ \ell_2$). 
In both cases the rescaling consists in dividing the
  counting function by the factor $N^\nu$, where
  $\nu=\dim(\cK)/2$. Reprinted from \cite{NoRu07}. \label{f:FWL-baker}}
\end{figure}

A strong form of fractal Weyl law was observed for an asymmetric
baker's map (see Fig.~\ref{f:FWL-baker}). For symmetric baker's maps 
(that is, taking $x_i=i/D$, see Fig~\ref{f:baker}), the fractal scaling seems
satisfied, but we observed that different profile functions occurred along
different geometric sequences $(N=N_o\,D^k)_{k\geq 0}$, a
manifestation of the number theoretic properties of such symmetric maps.
Apart from these specific number theoretic issues, the form of the
profile function for both the kicked rotor and the baker maps looks similar: $F(r)$
decays regularly from $r\approx 0$ and approximately vanishes around
some value $r_{\max}<1$, with
a ``dip'' before $r_{\max}$, showing a (mild) peak of the
density around some value $r_{peak}\leq r_{\max}$. The position of
this peak seems close to the classical decay
rate, $r_{peak}\approx e^{-\gamma_{cl}/2}$ \cite{Shep08}. 

A random matrix model Ansatz was proposed in \cite{schomerus} to account for the
profile function $F(r)$, hinting at a
certain ``universality'' of this profile, but the
validity of this Ansatz remains unclear.

\subsubsection{A solvable model satisfying the fractal Weyl law\label{s:Walsh}}
A ``toy-of-the-toy'' model was studied in \cite{NZ1,NZ12}, in the form
of a nonstandard quantization of the symmetric $D$-baker's map.
In the case of quantum dimension $N=D^k$, this quantization amounts to replacing the discrete
Fourier transform $F_N$ by the Walsh-Fourier
transform, that is the Fourier transform on $(\mathbb{Z}_D)^{k}$. This
quantization $M_{N}=M_{D^k}$
of the open baker's map then admits a very simple tensor product representation
on the Hilbert space $\mathcal{H}_N\equiv (\mathbb{C}^D)^{\otimes
  k}$. 

This property allows to {\it explicitly} compute the spectrum of
$M_N$: the latter is given in terms of
the $D\times D$ matrix $\Omega_D$, obtained by removing from the
inverse Fourier transform $F^*_D$ the $(D-n)$ columns corresponding to
the opening. Generally, this matrix has a $(D-n)$-dimensional kernel and
$n$ nontrivial eigenvalues, which are the eigenvalues of the $n\times
n$ square matrix $\tOmega_D$ extracted from $\Omega_D$.
This results in $n^k=N^{\nu}$ nontrivial eigenvalues (counted with
multiplicities) for $M_N$, and proves the strong form of 
 fractal Weyl law \eqref{e:FWL-maps}. The profile function $F(r)$ has the form of a
step function at some value $r_c=|\det\tOmega_D|^{1/n}$.

Yet, we noticed in \cite[Remark 5.2]{NZ1} that for some choices of parameters\footnote{It is
  the case, for instance, if we kill the second and fourth
  rectangles from the symmetric $4$-baker: in that case, $M_N$ has a
  single, simple nontrivial eigenvalue. One can cook up an even more
  dramatic example (with $D=16$, $n=2$), for which $\Spec(M_N)=\Spec(\tOmega_D)=\{0\}$.}, the spectrum
of $\tOmega_D$ may present an ``accidental'' extra kernel. In that case, the
counting function is $\cO(\hbar^{-\nu'})$ with $\nu'<\dim(\cK)/2$, so
even the weak fractal Weyl law fails. 
This accidental degeneracy seems
due to the very special tensor product structure of the Walsh
quantization, and should be nongeneric. 
It was checked \cite{Gendron09} that this accidental degeneracy disappears
if we modify the matrix $M_N$ by multiplying it by a diagonal matrix
of random, or even deterministic phases. Nevertheless, this problem may
indicate that any attempt to prove the fractal Weyl law in any setting might require some
{\it genericity} assumption, or the introduction of some random parameters in
the system.



\section{Interpretation of the fractal Weyl upper bound for open quantum
  maps\label{s:monodromy-fractal}}

After reviewing the numerical (and some analytical) results regarding
the optimality of the fractal Weyl upper bound, let us present
some heuristics for this upper bound in the case of open quantum
maps, as well as a rigorous proof for \emph{smooth} open quantum
maps. Both use a reduction of the dynamics to an effective propagator of
``minimal rank'', which accounts for the quantum dynamics near the
trapped set.

\subsection{Heuristic explanations}
A semiclassical mechanism explaining the fractal Weyl
upper bound for an open chaotic map $\kappa$ has been
put forward in \cite{schomerus}. The idea is that the (essential) generalized kernel of
$M(\hbar)$ is larger than its kernel (associated with the
opening), due to the presence of (approximate) Jordan blocks
reflecting the
transient classical dynamics of the points which wander through $V$ before
escaping.

\begin{figure}
\begin{center}
\includegraphics[width=0.8\textwidth]{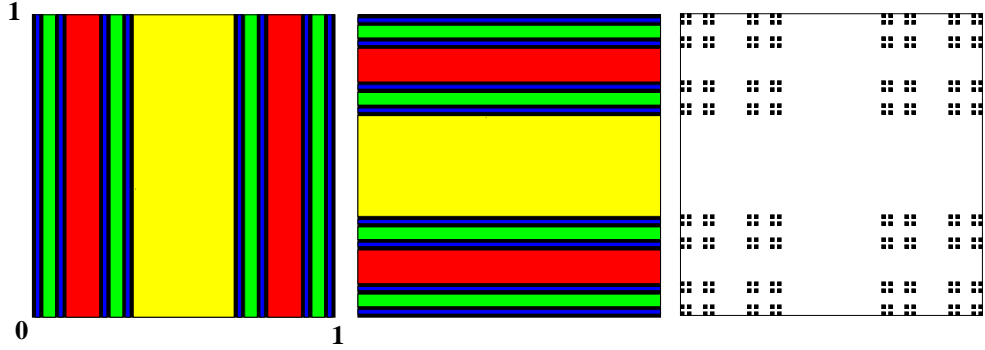}
\caption{Construction of the trapped set (right) and its incoming (left) and
  outgoing (center)
  tails for the symmetric $3$-baker. Each color corresponds to a
  specific escape time (in the futur or past) $n=1,\ldots,4$, the trapped
  set (and its tails) being approximated by union of black intervals/squares. \label{f:baker-trapped}}
\end{center}
\end{figure}
For any $n\geq 1$, consider the sets $D_n\in \TT^2$ of points
escaping before the time $n$. 
This set consists in a union of finitely
many connected components $D_{n,j}$, most of which look like ``thin tubes'' aligned along
the stable manifolds when $n\gg 1$ (see
Fig.~\ref{f:baker-trapped}, left). The widths of the thin tubes decay
like $e^{-n\lambda}$, where $\lambda$ is the
Lyapunov exponent. For fixed $n$, one can associate to each
component $D_{n,j}$ a quantum subspace $\cH_{\hbar,n,j}$ of dimension
$(2\pi \hbar)^{-1}\Vol(D_{n,j})$. The subspaces $\cH_{\hbar,n,j}$ are
semiclassically almost orthogonal
to each other, so that $\cH_{\hbar,n}=\bigoplus_j \cH_{\hbar,n,j}$
has dimension $(2\pi \hbar)^{-1}\Vol(D_{n})$.

The semiclassical evolution
implies that any state $u\in \cH_{\hbar,n}$ will
be absorbed when iterated up to time $n$:
$$
\|M(\hbar)^n u\| =\cO(\hbar^\infty) \|u\|,\qquad \forall u\in \cH_{\hbar,n}\,.
$$
This property implies that the long-living eigenvalues of
$M(\hbar)$ are essentially the same as those of
$M(\hbar)\rest_{\cH_{\hbar,n}^\perp}$, so their number is at most
$(2\pi\hbar)^{-1}(1-\Vol(D_{n}))$.

When $n\gg 1$ the set $\complement D_n$ of the points with escape times $>n$ is a small neighbourhood of
the incoming tail
$\cK^-$ (see Fig.~\ref{f:baker-trapped}), and
$$
\Vol(\complement D_n)\sim e^{-n\gamma_{cl}}\,,
$$ 
where $\gamma_{cl}>0$ is the classical decay rate.
To get a fractal upper bound, one needs to
push the time $n$ to infinity in a $\hbar$-dependent way.
As long as $n$ is smaller than the {\it Ehrenfest time} 
\be
T_{Ehr}\sim \frac{\log 1/\hbar}{\lambda_{\max}}\,,\quad\text{where $\lambda_{\max}$ is the largest expansion rate,}
\ee
the tubes $D_{n,j}$ have volumes $\gg\hbar$, which means that
one can associate nontrivial quantum subspaces $\cH_{\hbar,n,j}$ to
$D_{n,j}$. Ignoring problems due to the boundaries of $D_{n,j}$,
let us push the above argument up to $n=T_{Ehr}$:
the bound on the number of long-living eigenvalues then reads
$$
(2\pi\hbar)^{-1}\Vol(\complement D_{T_{Ehr}})\sim \hbar^{-1+\frac{\gamma_{cl}}{\lambda_{\max}}}\,.
$$
This argument is not optimal if the hyperbolicity is not homogeneous, the exponent
$1-\frac{\gamma_{cl}}{\lambda_{\max}}$ being larger than $\nu=\dim(\cK)/2$.
Still, the above reasoning clearly
exhibits the connection between resonance counting and small
($\hbar$-dependent) neighbourhoods of the trapped set $\cK$ (or its tail
$\cK^-$). It also shows the (approximate) Jordan structure of $M(\hbar)$,
quantum analogue of the transient dynamics before $T_{Ehr}$.
\begin{figure}[ht]
\begin{center}
\includegraphics[width=0.4\textwidth]{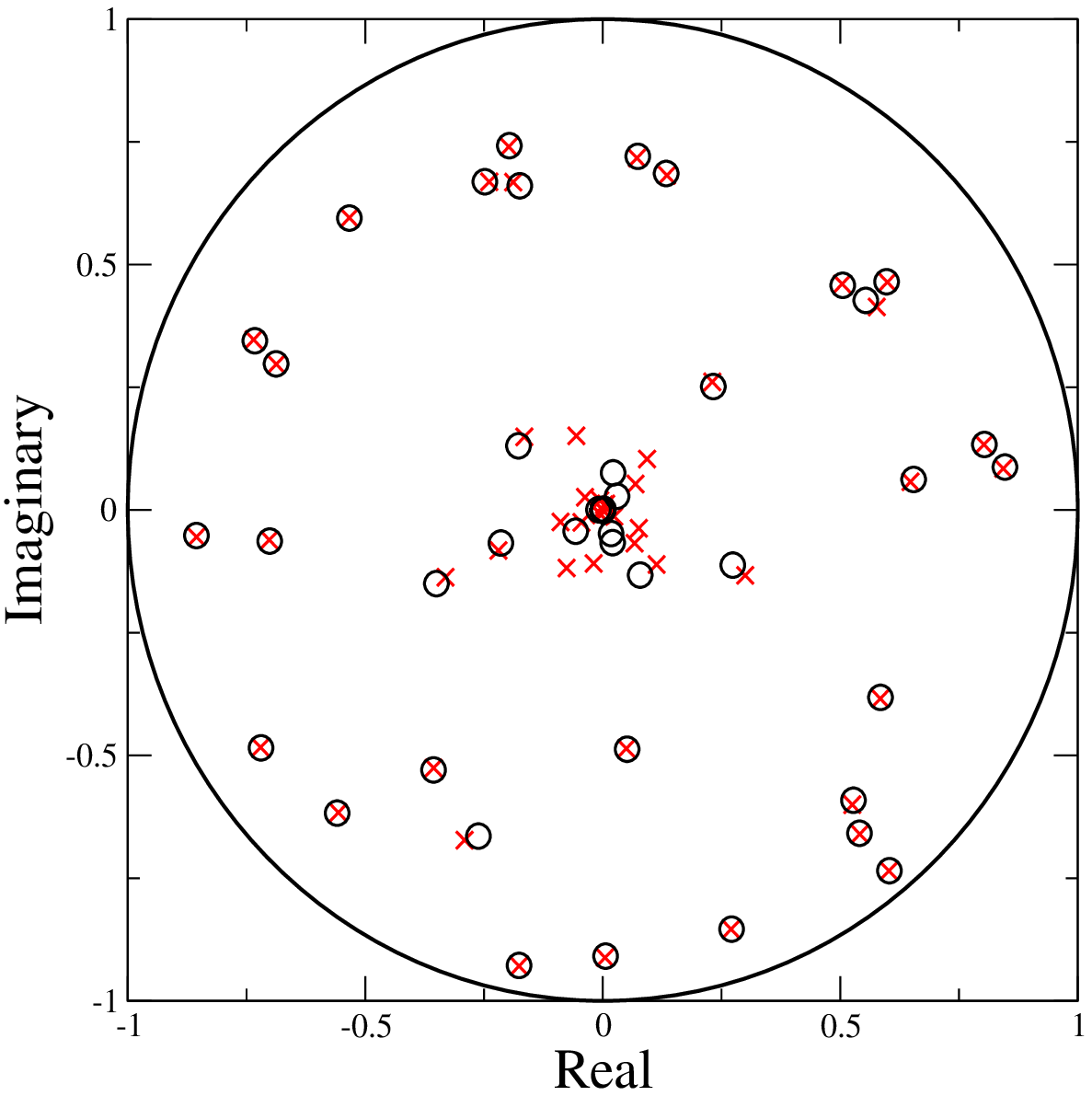}\hspace{1cm}
\includegraphics[width=0.5\textwidth]{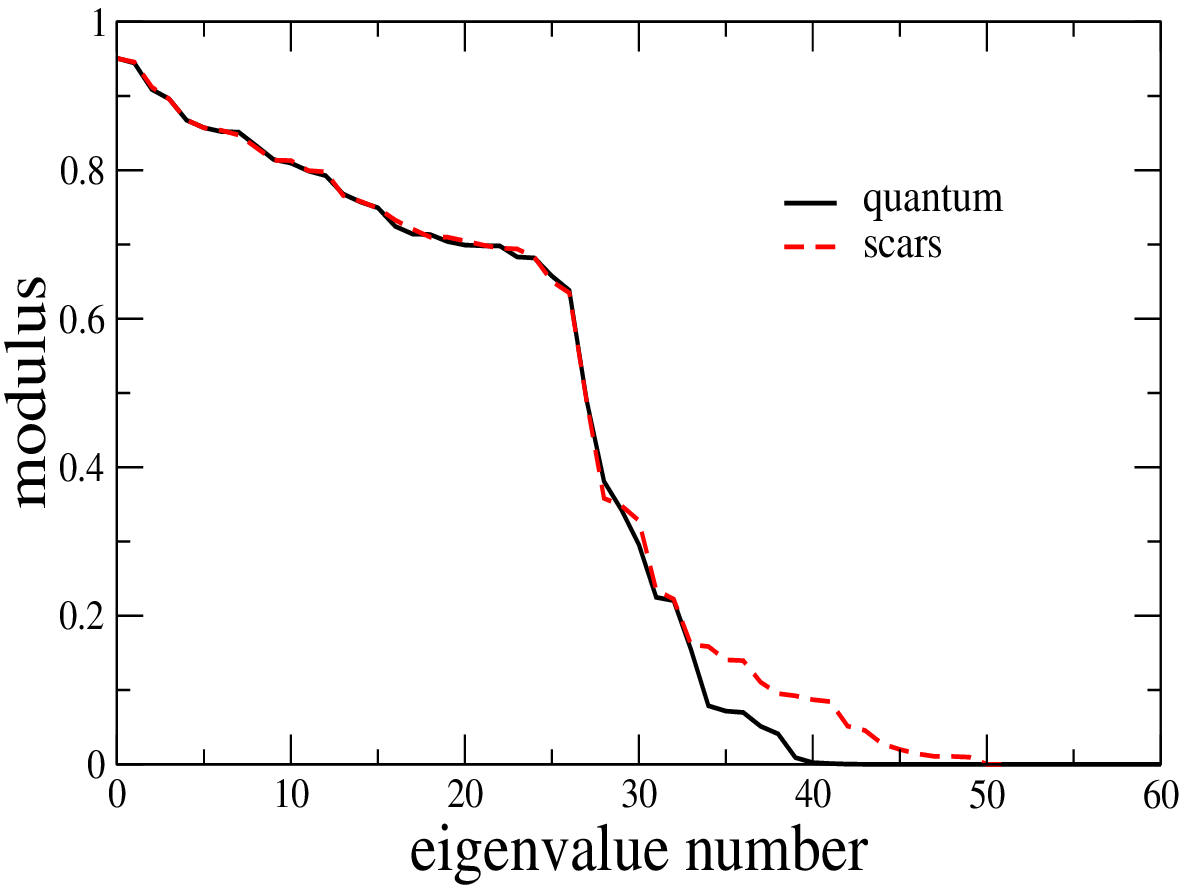}
\caption{Left: spectrum of the quantum open baker's map ($D=3$
  symmetric, $N=81$), by 
  diagonalizing the matrix $M_N$ (red crosses) or by using the ``scar
  matrix'' $E(z)$ from periodic orbits of length $\leq 5$, and solving \eqref{e:general} (black
  circles). Right: the corresponding radial countings (compared with
  Fig.~\ref{f:FWL-baker}, the two axes should be exchanged). 
Reprinted figure with permission from \href{http://link.aps.org/abstract/PRE/v80/e035202}{M. Novaes {\it et al.},
  Phys. Rev. {\bf E 80} 035202(R)  (2009)}. Copyright 2009 by the American Physical Society.\label{f:Novaes}}
\end{center}
\end{figure}
An alternative approach to the problem was adopted by Novaes {\it et
  al.} in \cite{Nov+09}. There the quantum dynamics was projected by
hand on a ``minimal'' quantum
subspace microlocalized near the trapped set,
resulting in an effective spectral problem of smaller dimension. The minimal
subspace was generated by a certain number ($\sim C\hbar^{-\nu}$) of
(Left and Right) {\it scar functions} $u^L_n$, $u_n^R$, that are
quasimodes of $M(\hbar)$, resp. $M(\hbar)^*$, microlocalized along
periodic orbits of periods $T\leq T_{Ehr}$.
The authors used these scar functions to construct the generalized eigenvalue problem
\be\label{e:general}
\det(E(\lambda)) = 0,\qquad E_{mn}(\lambda)\defeq\la u^L_n,(I -
\lambda^{-1}\,M(\hbar)) u_m^R\ra,
\ee
and checked that the solutions of this problem accurately
approximated the long living spectrum
of $M(\hbar)$ (see Fig.~\ref{f:Novaes}). 

The operator $E(\lambda)$, representing the
quantum dynamics on $\cK$, is an (approximate)
{\it effective propagator} for the quantum map $M(\hbar)$,
and can be considered to be of minimal rank, meaning that no further
reduction seems possible. Another advantage of this matrix
$E_{mn}(\lambda)$ is its sparsity: since quasimodes are localized near
periodic orbits, each quasimode $u_m^R$ (resp. iterated quasimode
$M(\hbar)u_m^R$) interferes only with
a few quasimodes $u_{n}^L$.

\subsection{A proof of the fractal Weyl upper bound\label{s:fractal-proof}}
A rigorous proof of the fractal upper bound for {\it smooth} open quantum
maps\footnote{see \S\ref{s:OQM}: the map $\kappa$ is smooth, and
the quasiprojector $\Pi(\hbar)=\Op(\alpha)$ is a ``nice'' pseudodifferential operator} can be
obtained using similar ideas. 
Notice that this smoothness condition excludes the open
quantum maps used in most numerical studies \cite{schomerus,NZ2,Shep08}.
\begin{thm}\cite{nsz2}\label{t:FWL-maps}
Consider $\kappa:V\mapsto \kappa(V)$ a (smooth) open map such that its trapped set
$\cK\Subset V$ is a hyperbolic repeller of upper Minkowski dimension $2\nu$, and a corresponding smooth
quantum map $\cM(\hbar)$, more generally a FIO $\cM(\alpha,\hbar)$
with symbol $\alpha\in C^\infty_c(V)$.

Then, for any $r>0$ and any small $\eps>0$, there exists $C_{r,\eps},\hbar_{r,\eps}>0$ s.t.
$$
\forall \hbar<\hbar_{r,\eps},\qquad \sharp \{\Spec(\cM(\hbar))\cap \{|\lambda|\geq r\}\}\leq
C_{r,\eps}\,\hbar^{-\nu-\eps}\,.
$$
If $\cK$ is of pure dimension, one can take $\eps=0$.\\
The same estimate holds for $M(\hbar)$ an open quantum map of finite rank.
\end{thm}
The proof is an ``exponentiation'' of the case of Schr\"odinger operators
presented in \S\ref{s:escape1}.
One constructs an escape function $G(x,\xi)$ on
$V\cup \kappa(V)$, such that 
\be\label{e:escape-M}
G\circ\kappa-G\geq 1\quad \text{outside an $\vareps$-neighbourhood $\cK^\vareps$ of
the trapped set $\cK$}.
\ee
To ensure that $G$ is a ``nice'' symbol, this
neighbourhood must have a width $\vareps\gtrsim \hbar^{1/2}$. 
One then quantizes this escape function into an operator $G^w$, and
uses the latter to conjugate the quantum map $\cM(\hbar)$
into 
$$
\cM_{tG}(\hbar) \defeq e^{-tG^w}\,\cM(\hbar)\,e^{tG^w}\,,\quad t\gg 1\,.
$$
The Egorov theorem and the pseudodifferential calculus show that
$\cM_{tG}(\hbar)$ is still a FIO, but with a
modified symbol $\alpha_{tG}\approx  \alpha\,e^{-t(G\circ \kappa - G)}$.
For $t\gg 1$ the escape property \eqref{e:escape-M} ensures
that $\cM_{tG}(\hbar)$ strongly suppresses the
states microlocalized outside $\cK^{\vareps}$.
One can thus construct a quantum
subspace $\cH_{\vareps}$ of dimension $\sim C\hbar^{-\nu}$
microlocalized on
$\cK^{\vareps}$, such that 
\be\label{e:H_eps}
\|(I - \Pi_{\vareps}) \cM_{tG}\|\ll 1\,,
\ee 
where $\Pi_{\vareps}$ is the orthogonal projector on $\cH_{\vareps}$.

From this remark, one can easily show that the number of long
living singular values of $\cM_{tG}$ is $\cO(\hbar^{-\nu})$, and get a
similar bound for its eigenvalues using Weyl's inequalities. 

We present an alternative argument, which has the advantage to apply as well to the
case of monodromy operators.
The property \eqref{e:H_eps} shows that, for any $\lambda\in\IC$ with $|\lambda|>r$,
the operator
\be\label{e:E(z)}
E(\lambda) \defeq (I - \lambda^{-1} \Pi_{\vareps} \cM_{tG}) -  \lambda^{-2}\,\Pi_{\vareps}
\cM_{tG}(I - \Pi_{\vareps}) \cM_{tG} \big[I-\lambda^{-1} (I - \Pi_{\vareps})
\cM_{tG} \big]^{-1}
\ee
is well-defined, and the second term on the RHS is a small correction
compared with the first one. Notice the similarity of the first term
with the operator in \eqref{e:general}.
A little algebra shows that the long living eigenvalues of $\cM_{tG}(\hbar)$ can
be exactly obtained by solving 
\be\label{e:detE(z)}
\det(E(\lambda))=0\,,\quad |\lambda|\geq r\,.
\ee
This confers to $E(\lambda)$ the role of an
{\it effective Hamiltonian} for the quantum map $\cM(\hbar)$. This operator ``minimally'' captures the long time
quantum evolution, which is ``supported'' on $\cK$. 
Applying Jensen's formula, one then
shows that the number of roots of \eqref{e:detE(z)} is
bounded from above by $\dim\cH_{\vareps}\sim C\hbar^{-\nu}$.

The same argument can be used to count the roots of $\det(I-M(z,\hbar))$,
with $M(z,\hbar)$ the quantum monodromy operators of Thm~\ref{thm:QMO}. One
defines an effective Hamiltonian $E(z)$ as in \eqref{e:E(z)}, replacing everywhere
$\lambda^{-1}M(\hbar)$ by $M(z,\hbar)$ \cite{nsz2}. 
This leads to a proof of Thms~\ref{thm:fractal}, and an alternative
proof for Thms.\ref{thm:fractal-h} and \ref{thm:fractal-Gamma} (under
the assumption that the trapped set $K_E$ is totally disconnected transversely to the flow).

\section{Resonance gap for open quantum maps and monodromy operators}\label{s:gap}
Let us now turn to Question (2), namely the criterion for a resonance gap
expressed in Theorems~\ref{t:gap-obstacles}
and \ref{t:gap-semiclass}. Below we will state the corresponding
result for open quantum maps and quantum monodromy
operators, which can also be used to prove these theorems. 

\begin{thm}[Spectral gap for open quantum maps]\label{t:gap-M}
Let  $\kappa:V\Subset\bSigma\mapsto \kappa(V)$ be a smooth open map
with hyperbolic trapped set $\cK$, and $\cM(\alpha,\hbar)$ an FIO
associated with $\kappa$ with symbol $\alpha\in C^\infty_c(V)$ nonzero
near $\cK$.

Then, for any small enough $\eps>0$, there exists $\hbar_\eps>0$ such
that the
spectral radii of the FIOs $\cM(\alpha,\hbar)$ satisfy
\be\label{e:gap-map}
\forall \hbar\leq \hbar_\eps,\qquad 
r_{sp}\big(\cM(\alpha,\hbar)\big)\leq \exp\big\{\cP(-\varphi^+/2 + \log|\alpha|,\kappa\rest_{\cK})+\eps\big\}\,.
\ee
Here $\varphi^+(\rho)\in C(\cK)$ is the logarithm of the unstable
Jacobian of $\kappa$, and $\cP(\bullet)$ is the topological
pressure. The case of open quantum maps corresponds to taking
$\alpha\equiv 1$ on $\cK$.

The same bound holds if we replace $\cM(\alpha,\hbar)$ by a finite
rank truncation $M(\alpha,\hbar)$ as in \S\ref{s:OQM}.
\end{thm}
We notice that the norm estimate \eqref{e:norm-est} implies the bound
\be\label{e:bound-norm}
r_{sp}(\cM(\alpha,\hbar))\leq \|\alpha\|_{\infty}+\cO(\hbar)\,.
\ee
This bound may be sharper than \eqref{e:gap-map},
depending on both $\kappa$ and $\alpha$. For instance, if $\alpha\equiv
1$ near $\cK$, then
\eqref{e:gap-map} is sharper than \eqref{e:bound-norm} iff the
pressure $\cP(-\varphi^+/2,\kappa\rest_{\cK})$ is negative, a condition
satisfied only provided $\cK$ is ``thin enough''.

Before sketching the proof of this theorem in the next section, let us explain how it can
be used to prove Thms~\ref{t:gap-obstacles}
and \ref{t:gap-semiclass}.
Eq.~\eqref{e:z-dependence} shows that the monodromy operator
$M(z,\hbar)$ associated with a scattering operator $P(\hbar)$ (or an
obstacle problem) has
the form of an FIO associated with a Poincar\'e return map $\kappa$, with symbol
$$
\alpha_z(\rho) = e^{-i\zeta\tau(\rho)} +
\cO(\hbar)\quad\text{near $\cK$}\,,\quad \zeta\defeq \frac{z-E}{\hbar}\,,
$$
where $\tau(\rho)$ is the return time. So, the relevant pressure
is $\cP(-\varphi^+/2 - \Im\zeta\tau,\kappa\rest_{\cK}\nolinebreak)$.

Let us assume that the root $s_0$ of the equation
\be\label{e:pressure-map}
\cP(-\varphi^+/2 -
s\tau,\kappa\rest_{\cK})=0\quad\text{satisfies }s_0<0\,.
\ee
Then, if we take $\Im \zeta \geq s_0+\tilde\eps$ for some $\tilde\eps>0$, the pressure 
$\cP(-\varphi^+/2 - \Im\zeta\tau)$ will be negative, and the
above theorem implies that, for $\hbar$ small enough, $r_{sp}(M(z,\hbar)) < 1$.
In turn, this bound implies (through Thm~\ref{thm:QMO}) that there are no resonance
in the strip $D(E,C\hbar)\cap\{\Im\zeta>s_0+\tilde\eps\}$. 
Finally, the theory of Axiom A flows \cite{BowRue75} shows that $s_0$ is
equal to the topological pressure of the flow:
$$
s_0=\cP(-\varphi_\Phi^+/2,\Phi^t\rest K_E)\,,
$$ 
where $\varphi^+_\Phi$ is the unstable Jacobian of the flow,
Eq.~\eqref{e:varphi+}. Hence, the condition $s_0<0$ is equivalent with the
conditions in Thms \ref{t:gap-obstacles} and \ref{t:gap-semiclass}.

\subsection{Proof of the resonance gap in terms of the topological pressure\label{s:proof}}
The proof of the spectral bound \eqref{e:gap-map} is analogous to the
case of the Schr\"odinger flow treated in \cite{NZ2}.
Let us assume that the trapped set $\cK$ is totally disconnected, which is
the case if we want to apply the result to monodromy operators. 
This restriction is not necessary, but it simplifies the proof a little.

To obtain an upper bound on the spectral radius of $\cM(\alpha,\hbar)$
(which we will denote by $\cM(\hbar)$ from now on), the
usual strategy is to estimate
$\|\cM(\hbar)^n\|$, with $n\gg 1$. Inspired by classical dynamical methods, we will proceed by
splitting $\cM(\hbar)^n$ into many components, each one being associated
with a ``pencil'' of classical trajectories. The topological
pressure will then naturally arise when summing over all the ``pencils''.

Let us be more precise. Using the assumption that $\cK$ is totally
disconnected, for any small $\eps>0$ we may consider 
a {\it Markov cover} $(V_a)_{a\in A_1}$ of the trapped set, such that
the open sets $V_a$ have diameters at most
$\eps$. The Markov property means the following: the sets $V_{a}$
are disjoint, and from them we may construct the transition matrix
$$
T_{a'a}=\begin{cases}1&,\quad V_a\cap
  \kappa^{-1}(V_{a'})\neq 0,\\ 0,&\quad\text{otherwise}\,.\end{cases} 
$$
Then, for any sequence of symbols
$\balpha=\alpha_{0}\alpha_1\cdots\alpha_{n-1}$, the set 
$$
V_{\balpha}\defeq V_{\alpha_0}\cap
\kappa^{-1}(V_{\alpha_0})\cap\cdots\cap\kappa^{-n+1}(V_{\alpha_{n-1}})
$$
is nonempty if and only if, at all steps $j=0,\ldots,n-2$, one has $T_{\alpha_{j+1}\alpha_j}=1$.
The set $V_{\balpha}$ consists of the initial points $\rho$ which
share the same ``symbolic history'' for times $0\leq j\leq n-1$; it
makes up a ``pencil'' of trajectories.

\begin{figure}[ht]
\begin{center}
\includegraphics[width=0.4\textwidth]{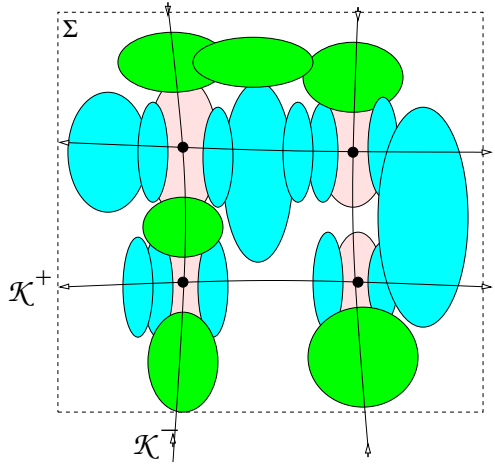}
\caption{Sketch of the open cover $(V_{a})_{a\in A_1}$ of the trapped
  set (pink), completed by the open sets $(V_{a})_{a\in A_+}$
  (cyan) and  $(V_a)_{a\in A_-}$ (green) away from $\cK$. Black lines
  indicate the tails $\cK^{\pm}$ and black circles the trapped set. \label{f:cover-map}}
\end{center}
\end{figure}

To each set $V_a$ we associate the weight 
\be\label{e:w_a}
w_{a}=\max_{\rho\in \cK\cap V_a}\,e^{-\varphi^+(\rho)/2}\, |\alpha(\rho)|\,,
\ee
and consider the weighted transition matrix
$T^w_{a'a}=T_{a'a}\,w_{a}$. 
The topological pressure appearing in \eqref{e:gap-map} is then approximated by the largest
(Perron-Frobenius) eigenvalue of the matrix $T^w$:
\be\label{e:pressure-PF}
\cP(-\varphi^+/2+\log|\alpha|,\kappa\rest_{\cK})=\lim_{\eps\to 0}\log\lambda_{PF}(T^w)\,.
\ee
We may complete this Markov cover into an open cover of $V$,
\be\label{e:cover}
V\subset\cup_{a\in A}V_a\,,\quad A=A_1\cup A_+\cup A_-\,,
\ee
such that, for some time $n_o$, all sets $V_{a_-}$, $a_-\in A_-$
(resp. $V_{a_+}$, $a_+\in A_+$)
escape in the hole before the time $n_o$ in the backwards (resp. forward)
evolution, see Fig.~\ref{f:cover-map}. 
To the cover $(V_a)$ we associate a smooth partition
of unity of the phase space $\bSigma$, namely a finite collection of
cutoffs $\chi_a\in C^\infty_c(V_a,[0,1])$, satisfying
$$
\sum_{a\in A}\chi_a\equiv 1 \quad\text{in some neighbourhood of $V$},
$$
and add a component in the hole, $\chi_{\infty}=1-\sum_{a\in A}\chi_a$
to get a full partition of unity.
This smooth partition of unity is then quantized into a quantum
partition 
$$
Id =\sum_{a\in A\cup\infty}\Op(\chi_a)\,,
$$
which is used to split the iterated propagator $\cM(\hbar)^n$ into
components: 
$$
\cM(\hbar)^n =
\sum_{|\balpha|=n} \cM_{\balpha},\qquad \cM_{\balpha}=\cM_{\alpha_{n-1}}\,\cM_{\alpha_{n-2}}\cdots
\cM_{\alpha_0},\qquad \cM_{a}\defeq \cM(\hbar)\,\Op(\chi_a)\,.
$$

\subsubsection{Analyzing the components $\cM_{\balpha}$}
The advantage of this decomposition is to obtain an upper
bound for the individual components $\cM_{\balpha}$ which is sharper
than the obvious bound
\be\label{e:bound-trivial}
\|\cM_{\balpha}\|\leq (\|\alpha\|_{\infty}+\cO(\hbar))^n\,.
\ee
For this, we use our knowledge of the classical dynamics. 
From Egorov's theorem, we know that for any finite sequence $\balpha$,
the element $\cM_{\balpha}=\cO(\hbar^\infty)$ unless the set
$V_{\balpha}\neq\emptyset$; the sequence
$\balpha$ is then called {\it admissible}. Also, any sequence
containing $a=\infty$ leads to a negligible term.

As a result, since the sets
$V_{a_-}$ (resp. $V_{a_+}$) escape before the time $n_0$ in the past (resp. in the future),
we deduce that for any $n>2n_0$ the only nonnegligible components
$\cM_{\balpha}$ must be of the form 
$$
\balpha = \balpha_+\balpha^{(1)}\balpha_-,\qquad |\balpha_-| = |\balpha_+|=n_0,\quad
\balpha^{(1)}\in A_1^{n-2n_0}\quad \text{admissible}.
$$
In view of this property, for $n\gg 1$ we may restrict ourselves to
the admissible sequences $\balpha\in A_1^n$, that is replace
$\cM(\hbar)$ by$\cM_{A_1}\defeq \sum_{a\in A_1}\cM_a$.

\subsubsection{Acting on a Lagrangian state: a hyperbolic dispersive estimate}

We now want to use the hyperbolicity of $\kappa$ near $\cK$. 
If we apply the FIO $\cM_{\alpha_0}$ to a Lagrangian\footnote{If the
  Lagrangian $\Lambda=\{(x,dS(x))\}$ for the generating function $S(x)$,
 a Lagrangian (or WKB) state associated with $\Lambda$ has the form $u_0(x) = f(x)\,e^{iS(x)/\hbar}$, with $f\in
C^\infty_c$.}
state $u_0$ supported by a Lagrangian manifold $\Lambda_{\alpha_0}$ transverse
to the stable direction
$E^-$, the state will expand along the unstable direction. If the resulting
state spreads outside $V_{\alpha_1}$, cutting it through
$\Op(\chi_{\alpha_1})$ will reduce
its norm by a finite factor, while the output state will again be a WKB state along a Lagrangian
$\Lambda_{\alpha_1\alpha_0}$ transverse to $E^-$. This phenomenon repeats itself, and
leads to the following 
{\it hyperbolic dispersive estimate}:
\be\label{e:HDE}
\|\cM_{\balpha}u_0\| \leq C\, w_{\balpha},\qquad w_{\balpha}=\prod_{j=0}^{n-1}w_{\alpha_j}\,.
\ee
Because the unstable Jacobian is
bounded below, $\varphi^+(\rho)\geq\Lambda^+>0$, the weights satisfy 
$$
w_a\leq e^{-\Lambda^+}\|\alpha\|_\infty\,,
$$
so after a some time the estimate \eqref{e:HDE} becomes sharper than
the bound \eqref{e:bound-trivial}.

This type of estimate first appeared in
the work of Anantharaman on Anosov flows \cite{Anan08}. It was
extended to the case of scattering problems with a hyperbolic
repeller in \cite[Prop.6.3]{NZ2} (see also \cite[Section 4]{CRM10}), and to
situations with a nonconstant symbol $\alpha$ in \cite{Schenck10}.

\subsubsection{Putting the pieces together}
One applies the triangular inequality to get a bound on the sum
of terms made up by $\cM_{A_1}^n$:
$$
\|\cM^n_{A_1}u_0 \| \lesssim 
\sum_{\balpha\in A_1^n\ admis.} w_{\balpha} = \sum_{a',a\in A_1} [(T^w)^n]_{a' a}\,,
$$
where we use the weighted transition matrix.
For $n\gg 1$, the high power of the matrix $T^w$ is dominated by its Perron-Frobenius
eigenvalue, which, according to \eqref{e:pressure-PF}, is close to the topological pressure,
so for some $\tilde\epsilon>0$ we get, 
\be\label{e:M^ne}
\|\cM^n_{A_1}u_0 \|\leq C\,e^{n(\cP(-\varphi^+/2+\log|\alpha|)+\tilde\eps)}\,.
\ee
So far we have considered the action of the propagator on very particular
Lagrangian states $u_0$.
However, any state $u$ microlocalized in $V_{\alpha_0}$ can be
expanded into a ``basis''  $(u_{\zeta})_{\zeta\in
W}$ of such WKB states:
\be\label{e:decompo}
u = \int_{W} \frac{d\zeta}{(2\pi\hbar)^{d/2}}\,
\hat{u}(\zeta)\,u_\zeta + \cO(\hbar^\infty)\,,
\ee
with $W$ a bounded domain in $\IR^{d}$ and $\int_W d\zeta
|\hat{u}(\zeta)|=\cO(1)$. 
Applying $\cM_{A_1}^n$ to the decomposition
\eqref{e:decompo} and adding the contributions of the ``tails''
$\balpha_{\pm}$, we obtain for $n\gg 1$ the norm
estimate 
$$
\| \cM(\hbar)^{n}\|\leq C\,\hbar^{-d/2} \,e^{n(\cP(-\varphi^+/2+\log|\alpha|)+\tilde\eps)}\,.
$$
Crucially, the above estimate is valid for ``large logarithmic times''
$n\sim \tilde\eps^{-1}\log (1/\hbar)$, for which we have
$\hbar^{-d/2}\leq e^{n\tilde\eps}$. We thus get
\be\label{e:Mn}
\| \cM(\hbar)^{n}\|\leq \exp\{n \big(\cP(-\varphi^+/2+\log|\alpha|) + 2\tilde\eps\big) \}\,,
\ee
which proves the spectral bound \eqref{e:gap-map}.
\qed

\subsection{Is the pressure bound optimal?}\label{s:pressure-optimal?}

As shown above, the pressure bound  \eqref{e:Mn} is obtained by evolving
Lagrangian states $u_0$ through the
components $\cM_{\balpha}$, resulting in the hyperbolic dispersive
estimate \eqref{e:HDE}, which is generally sharp. Then we applied the triangular inequality
to bound the norm of $\cM^n_{A_1}u_0$, and got the bound \eqref{e:M^ne}.
The question is: how much does one ``lose'' through this
triangular inequality? 

The square norm of $\cM^n_{A_1}u_0$ can be written as 
$$
\la \sum_{\balpha\in A_1^n\ admis.} \cM_{\balpha}u_0,\sum_{\balpha\in A_1^n\ admis.} \cM_{\balpha}u_0\ra\,.
$$
If the states $\cM_{\balpha}u_0$ were orthogonal to
each other, this scalar product would be given by a diagonal sum
\be\label{e:pressure-1}
\sum_{\alpha\in A_1^n\ admis.} \|\cM_{\balpha}u_0 \|^2 \leq C\sum_{\balpha\in A_1^n\ admis.}
w_{\balpha}^2\leq \exp\big\{n \big(\cP(-\varphi^+ +2\log|\alpha| ) + \tilde\eps\big)\big\}\,.
\ee
The bound is sharper than \eqref{e:M^ne}, because for any nonzero
test function $f$ one has $\cP(2f)<2\cP(f)$. 
For instance, in the case of an open quantum map, $\alpha\equiv 1$ near $\cK$, the
pressure $\cP(-\varphi^+)=-\gamma_{cl}$ is always negative, while
$\cP(-\varphi^+/2)$ is negative only provided $\cK$ is ``thin enough''.

Two states $\cM_{\balpha}u_0$, $\cM_{\balpha'}u_0$
will indeed be (essentially) orthogonal if the final indices
$\alpha_{n-1}\neq\alpha_{n-1}'$ (the states are localized
in disjoint sets), or if they are supported on Lagrangian leaves
$\Lambda_{\balpha},\Lambda_{\balpha'}\subset V_{\alpha_{n-1}}$
at distance $\gg \hbar$
from one another: a nonstationary phase estimate then ensures that
$$
\la \cM_{\balpha'}u_0, \cM_{\balpha}u_0\ra =  \cO(\hbar^\infty)\,.
$$
This essential orthogonality indeed occurs for sequences of length $n\leq c\log(1/\hbar)$, with
$c>0$ sufficiently small.
But for the large logarithmic times
$n\sim\tilde\eps^{-1}\log(1/\hbar)$ we need, many states $\cM_{\balpha}u_0$ will be
supported by Lagrangians $\hbar$-close to one another, leading to
nonnegligible off-diagonal terms
$$
\la\cM_{\balpha}u_0,\cM_{\balpha'}u_0\ra\approx
e^{i(\theta_{\balpha}-\theta_{\balpha'})} \,w_{\balpha}\,w_{\balpha'}\,.
$$
The phases $\theta_{\balpha}$, $\theta_{\balpha'}$ are the actions
accumulated along the ``paths'' $\balpha$, $\balpha'$;
it is tempting to believe that these phases are {\it
  pseudo-random}. Namely, that they behave like
\emph{independent} random phases: the sum of the off-diagonal elements
would then be of the {\it same} order as the sum \eqref{e:pressure-1} over diagonal terms, and
lead to a spectral bound 
\be\label{e:optimal?}
r_{sp}(\cM(\alpha,\hbar))\leq e^{\cP(-\varphi^+ +2\log|\alpha| )/2+\tilde\eps}\,.
\ee

\subsubsection{Phase cancellations in classical dynamics}
Even if true, the pseudo-randomness of the phases $\theta_{\balpha}$ seems very difficult to prove.
What can be done rigorously? {\it Partial} phase cancellations were exhibited
by Dolgopyat in his proof of exponential mixing for contact Anosov flows
\cite{Dol98,Live04}. 
In this situation the FIOs $\cM(\alpha,\hbar)$ are replaced by Ruelle's transfer
operator $\cL_s$ associated with a certain expanding map $T$, defined on some
unstable leaf $W^+$ by  projecting $\kappa$ along the stable foliation:
$$
\cL_{s}\,u (x) \defeq \sum_{y:T(y)=x} e^{-s\tau(y)}\,u(y)\,,\quad
\text{so}\quad \cL_{s}^n\,u (x) \defeq \sum_{y:T^n(y)=x} e^{-s\tau_n(y)}\,u(y)
$$
Here $\tau$ is the Poincar\'e return time, and $\tau_n$ is the time
accumulated after $n$ iterations, and the
parameter $s=s_0+it$, where the imaginary part $t$ should be compared
with $\hbar^{-1}$. Using a nonintegrability property of the return
time, Dolgopyat showed that for $t$ large enough
partial phase cancellations occur in the above sum for $\cL_{s}^nu$,
leading to a shrinking of the spectral radius:
\be\label{e:Dolgo}
\exists \eps_0>0,\,t_0>0,\ \forall t\geq t_0,\qquad r_{sp}(\cL_{s_0+it})\leq r_{sp}(\cL_{s_0})\,e^{-\eps_0}\,.
\ee
Unfortunately, the improvement $\eps_0$ is hardly explicit.

\medskip

A similar improvement was obtained in the case of the Laplacian on convex co-compact hyperbolic surfaces
$X=\Gamma\backslash \IH^2$ (see \S\ref{s:geom}).
Following \cite{JakNaud10}, let us define the {\it essential spectral gap} representing the optimal
resonance free strip at high frequency:
$$
G(X)\defeq \inf \{\sigma\leq n/2,\ \Res(\Delta_X)\cap \{\Re s\geq \sigma\}\
\text{is finite}\}\,.
$$
Here the ``pressure'' bound \eqref{e:X-gap} means that
$G(X)\leq \delta$. 
Using the characterization of the resonances in terms of a certain Ruelle
transfer operator $\cL_s$, Naud \cite{Naud05} proved an improved
spectral bound of the form \eqref{e:Dolgo} and showed that the pressure bound
on $G(X)$ could be improved:
$$
\exists\eps_1>0,\quad G(X)\leq \delta-\eps_1\,.
$$
Jakobson and Naud further investigated the location of
resonances for certain {\it arithmetic} convex co-compact surfaces,
in both cases of  ``thick'' ($\delta\in [1/2,1]$) and ``thin'' ($\delta\in
(0,1/2)$) trapped sets \cite{JakNaud10}. Analyzing Selberg's zeta function, they managed to
prove {\it lower bounds} for the essential spectral gap:
\be
\text{"thick" $K$:}\quad G(X)\geq \frac{\delta}{2}-\frac14\,,\qquad
\text{"thin" $K$:}\quad G(X)\geq \frac{\delta(1-2\delta)}{2}\,.
\ee
They also conjecture that in both cases the essential
spectral gap should be given by
\be\label{e:conj-G}
G(X)=\frac{\delta}{2}\,.
\ee
This conjecture is equivalent with the value $\cP(-\varphi^+)/2=-\gamma_{cl}/2$
appearing in \eqref{e:optimal?} (in the case $\alpha\equiv 1$). Therefore, the conjecture
\eqref{e:conj-G} amounts to assume that the
phases appearing in $\cL_s^nu$
cancel each other at least as much as if they were random.
 
This conjecture is inspired by the case of arithmetic surfaces of
{\it finite} volume (e.g. the modular group $\Gamma=SL(2,\IZ)$), for which
the high frequency resonances are actually eigenvalues embedded in the absolute
spectrum, with $\Re s_j = 1/2 = \delta/2$. 

\medskip

A Dolgopyat type estimate \eqref{e:Dolgo} was also shown by Stoyanov in the case of
classical scattering by $J$ convex obstacles in two dimensions \cite{Stoy01}
or higher dimensions \cite{Stoy10} and also for
more general Axiom A flows \cite{Stoy11}. In the case of scattering by
convex obstables, there is no exact connection between the zeros of
the semiclassical (Selberg-type) zeta function and the quantum
resonances. Yet, Petkov and Stoyanov \cite{PetStoy10} were able
to compare the long time quantum evolution
$\cM(k)^n u_0$ for some initial Lagrangian state $u_0$, with a
(modifed) evolution of $u_0$ through a {\it classical} transfer
operator of the form $\cL_{s}$, $s=ik$. This connection allowed
them to use the improved spectral gap for $\cL_s$, $|\Im s|\gg 1$, to (effectively)
get a smaller spectral radius for $\cM(k)$ than predicted in \eqref{e:gap-map}, hence a wider resonance
free strip than predicted in Thm~\ref{t:gap-obstacles}. 

How large could the resonance gap be for such obstacle problems? Could
it be as large as $\gamma_{cl}/2$, as conjectured above for hyperbolic surfaces?
As noticed in \S\ref{e:sharp-FWL?}, the
numerics performed for the $3$-disk scattering on $\IR^2$
\cite{LuSrZw03} shows a peak in the resonance density centered near $\Im
k=-\gamma_{cl}/2$. These numerics are unable to predict how the peak
behaves in the high frequency limit. 
If this peak remains of {\it positive width} when $k\to\infty$,
this would indicate that the resonance gap is 
smaller than $\gamma_{cl}/2$.

It is very likely that this improvement on the resonance gap can be extended to the case of
semiclassical Schr\"odinger operators \eqref{e:Hamiltonian-V} with
hyperbolic repellers; the main difficulty probably resides in checking
that the classical conditions for a Dolgopyat estimate to hold are met. 

\subsubsection{Some numerics for the open baker's map\label{s:radius-baker}}
Most numerical studies of open quantum maps were focussing on
the spectral density and the fractal Weyl law, rather than the spectral
radii. We have mentioned in \S\ref{s:maps-fractalWeyl}
that the numerics relative to several open maps
show a peak in the radial spectral distribution near a value
$r_{peak}\approx e^{-\gamma_{cl}/2}$, which
``pushes'' the spectral radius to a larger value.
Below we provide some numerical results for the
open baker's map (see \S\ref{s:baker}).

We only consider symmetric baker's map with $D$
symbols (that is the map \eqref{e:baker} with $x_i=i/D$), so
that the unstable Jacobian $\varphi^+\equiv\log D$.
We let the hole consist in the union of $D-n$ Markov rectangles
($0<n<D$).
The topological pressure is then given by
$$
\cP(-s\varphi^+) = \log n - s\log D\,.
$$
In particular, this pressure does not depend on {\it which} rectangles
are removed, but only on their number. The baker's map is
discontinuous along the boundaries of the rectangles, and these
discontinuities are believed to induce diffraction effects at the
quantum level; for
this reason, the open baker's map does not satisfy the assumptions of
Thm~\ref{t:gap-M}. Yet, if the
leftmost and rightmost rectangles both belong to the hole, the trapped set $\cK$ is
at finite distance from the discontinuities, so it is reasonable to
expect that the quantum spectrum should not be too sensitive to this diffraction.
\begin{figure}
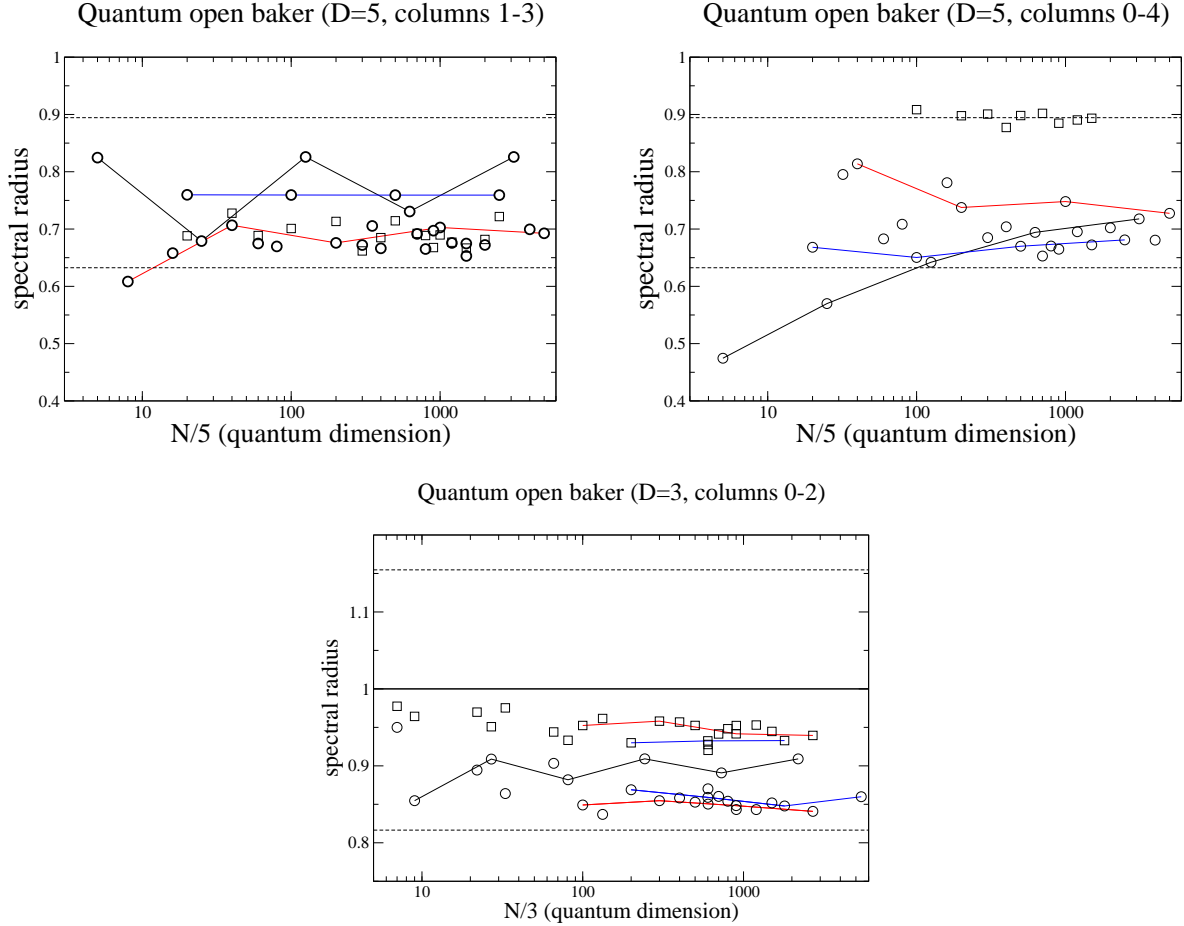

\begin{center}
\includegraphics[width=0.45\textwidth]{baker5-gap.eps}\hspace{1cm}\includegraphics[width=0.45\textwidth]{baker5b-gap.eps}\\
\vspace{.4cm}
\includegraphics[width=0.45\textwidth]{baker3-gap.eps}
\caption{Spectral radii of various quantum symmetric open baker's
maps, with $n=2$ kept rectangles. $N=(2\pi\hbar)^{-1}$ is the quantum
dimension. Eigenmodes are split according to parity:
even ($\circ$) vs. odd ($\square$). Straight lines
emphasize geometric series $N=N_oD^k$, $k=0,1,2,\ldots$. 
Top left: $D=5$, kept rectangles $i=1,3$. Top right: $D=5$, kept
rectangles $i=0,4$. 
Bottom: $D=3$, kept rectangle $i=0,2$. The dashed horizontal lines indicate the values
$e^{\cP(-\varphi^+/2)}>e^{\cP(-\varphi^+)/2}$. 
\label{f:baker-gap}}
\end{center}
\end{figure}
We have numerially computed the spectral radii of several open quantum
baker's maps (see Fig.~\ref{f:baker-gap}). The quantum dimension
$N=(2\pi\hbar)^{-1}$ was taken in a range $10\lesssim N \lesssim
5000$. The first case (top left) is a baker with $D=5$
symbols, where the kept rectangles have indices
$i=1,3$, so that the trapped set $\cK$ is away from the
boundaries. The spectral radii $r(N)$ seem to satisfy 
$$
e^{\cP(-\varphi^+)/2} +\eps_1 < r(N) < e^{\cP(-\varphi^+/2)}-\eps_2\,,
$$
for some $\eps_i>0$, but keep fluctuating for large $N$. 
The parity of the eigenmodes (w.r.t. $x=0$) does not seem to play an
important role.

On the opposite, in the case of the $D=5$ baker with kept rectangles
$i=0,4$, the trapped set contains
the axes $\{x=0\}$, $\{\xi=0\}$, on which the map is discontinuous. The spectral radii of the even-parity sector
satisfy the same bound as above, but the odd-parity
spectral radius $r_{odd}(N)\approx e^{\cP(-\varphi^+/2)}$, thus barely
violating the bound of Thm~\ref{t:gap-M}. The reason why the spectral
radius for the odd sector is larger than for the even one is unclear.

In the case $D=3$ (bottom), the trapped set also touches the
discontinuity set;
the value $e^{\cP(-\varphi^+/2)}>1$ is larger than
the unitarity bound. As above, odd states show larger eigenvalues than even
ones. The radii seem to satisfy 
$$
e^{-\cP(-\varphi^+)/2} +\eps_1 < r(N) < 1-\eps_2\,,\qquad \eps_i>0\quad\text{fixed}\,,
$$
indicating a gap below the unitarity bound.

On the figures we emphasize some geometric series $N=N_oD^k$,
$k=0,1,2,\ldots$, because such series were shown important when
studying the fractal Weyl law \cite{NZ1,NZ12}. The spectral radii along such series indeed
show some regularity, especially when taking the parity of $k$ into account.

\bigskip

In the case of the Walsh quantization of the symmetric $D$-baker's map presented in
\S\ref{s:Walsh}, the spectral radius of $M_N$ (for $N=D^k$, $k\in\IN$)
is given by the largest eigenvalue $\lambda$ of the $n\times n$ matrix
$\tOmega_D$, obtained by removing from the $D$-discrete
Fourier transform the $D-n$ columns and lines corresponding to the
hole. As a result, the spectral radius of $M_N$ is the same for all quantum
dimension $N=D^k$. The pressure bound 
$$
r_{sp}(M_N)\leq e^{\cP(-\varphi^+/2)}=\frac{n}{\sqrt{D}}
$$
obviously results from the fact that all entries of
$\tOmega_D$ have modulus $\frac{1}{\sqrt{D}}$.
The various examples we have treated in \cite{NZ1,NZ12} show that for
this model the
spectral radius is unrelated with the value
$e^{-\gamma_{cl}/2}=\sqrt{\frac{n}{D}}$, its values can vary across the
full interval $[0,\min(1,\frac{n}{\sqrt{D}})]$ (including the
extremal values), depending on
the explicit phases in the matrix $\tOmega_D$.
This situation is far from generic, and seems to rely on the fact that the underlying harmonic analysis is associated with the
Walsh-Fourier transform.

A preliminary investigation of a smoothed version of the
(standard) quantum baker's map apparently leads to a Dolgopyat type
partial phase cancellation, which would then force the spectral radius
to be $\leq \frac{n}{\sqrt{D}}-\eps_2$ for some $\eps_2>0$. Yet, in
  spite of the explicit (and relatively simple) expressions for the
  the phases, it seems impossible to push these cancellations such as
  to recover the conjectured ``optimal'' bound $\sqrt{\frac{n}{D}}$.


\section{Phase space structure of wavefunctions\label{s:metastable}}
In this last section we address Question (3), that is we investigate the structure of the
``eigenfunctions'' (in a generalized sense) of our scattering system introduced in
\S\ref{s:resonances}.

The first class of such eigenfunctions will be the metastable states $u_j(\hbar)$ associated with the
(discrete) resonances $z_j(\hbar)$. They satisfy the differential
equation $(P(\hbar)-z_j)u_j=0$, are purey outgoing and blow up
exponentially at infinity.

On the opposite,  for any real energy $E$ the 
scattering functions form an infinite
dimensional space of functions satisfying $(P(\hbar)-E) u=0$. They are not
square-integrable either, but contain both incoming and outgoing components.

In both cases, we will focus on the structure of these functions in
the interaction region, say the ball $B(0,R_0)$.

\subsection{Metastable states\label{s:meta}}
Let $u_j(\hbar)$ be the metastable state associated with a resonance
$z_j(\hbar)$ of our scattering Hamiltonian $P(\hbar)$. We may
(somewhat arbitrarily) normalize this state inside the interaction
region $B(R_0)$, by putting
$$
\|u_j(\hbar)\|_{L^2(B(R_0))}=1\,.
$$
In order to connect
oneself with the classical dynamics, it is natural to analyze the modes $u_j(\hbar)$
in terms of their associated {\it phase space distributions}. Let us
recall that to any
function $u\in L^2$ one can associate its Wigner function
$W_u^\hbar(x,\xi)$, depending quadracally on $u$ (the formula is given
in \eqref{e:Wigner}). This Wigner function (or the Husimi
function obtained by a smoothing on the scale $\sqrt{\hbar}$) is
interpreted\footnote{The Wigner function generally takes both positive
and negative values, which makes this interpretation a bit
questionable. On the opposite, the Husimi function is nonnegative.} as a phase
space probability density for the state $u$. The distribution
$W^\hbar_u=W^\hbar_u(x,\xi)dx\,d\xi$ will be called the Wigner
distribution (or signed measure). 

Describing the individual
functions $W^\hbar_{u_j(\hbar)}$ seems a hopeless task. On the other
hand, it is often possible to derive some asymptotic properties of a
\emph{semiclassical sequence} of such functions. 

Take any sequence $\hbar_k\to 0$, and for each
$\hbar=\hbar_k$ choose some resonance $z(\hbar)=z_j(\hbar)\in
D(E,C\hbar)$, a corresponding 
metastable state $u(\hbar)=u_j(\hbar)$, and construct its Wigner
distribution $W^\hbar_{u(\hbar)}$. We will then be interested in the
asymptotic behaviour of the sequence $(W^\hbar_{u(\hbar)})_{\hbar=\hbar_k}$ when
$\hbar\to 0$.

Bony and Michel \cite[Thm~2.1]{BonyLaurent03} showed
that, for a general trapping potential, the Wigner distributions
$W_{u(\hbar)}$ are semiclassically negligible away from the
outgoing tail $K^+_E$: for any test function $a\in C^\infty_c(T^*X)$
supported away from $K_E^+$, one has 
$$
\la W^\hbar_{u(\hbar)},a\ra=\cO(\hbar^\infty)\,.
$$
This estimate does not depend on the structure of the flow on $K_E$.

\begin{figure}[ht]
\begin{center}
\includegraphics[angle=-90,width=1.\textwidth]{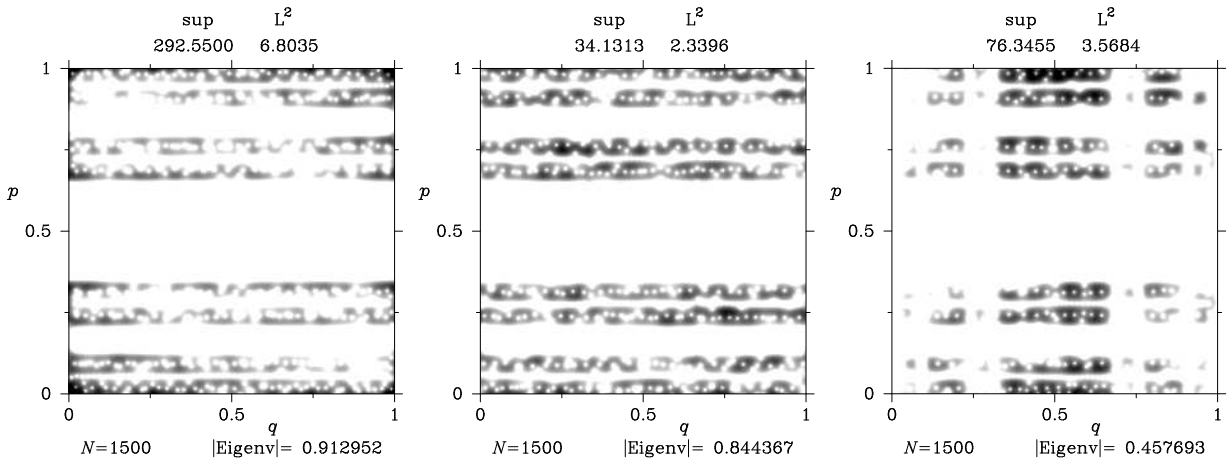}
\caption{Husimi functions of three metastable states of the quantum
symmetric open 3-baker (logarithmic grey scale). The high intensities (black) are clearly
  localized on $\cK^+$.\label{f:Husimi}}
\end{center}
\end{figure}

The same type of result was also obtained in the case of the open
baker's map \cite{Keating+06,NoRu07}, where some Husimi measures
were numerically computed, and shown to
concentrate on the outgoing tail $\cK^+$ (see this set on
Fig.~\ref{f:baker-trapped}, center, and compare with
Fig.~\ref{f:Husimi}). 

A more precise asymptotic description of the Wigner (or Husimi)
distributions is provided by the concept of \emph{semiclassical
  measure}, that is a measure $\mu$ on phase
space\footnote{We will only consider the restriction of this measure on the
  interaction region $T^*B(0,R_0)$.}, obtained as a 
limit (in the weak-$*$ topology) of the sequence of distributions
$(W^\hbar_{u(\hbar)})_{\hbar\to 0}$, equivalently \emph{the} limit of 
certain extracted
subsequence $(W^\hbar_{u(\hbar)})_{\hbar\in S}$  ($S$ is some infinite
subsequence of $(\hbar_k)$). 
This measure describes the asymptotic phase space distribution of the
metastable states along the subsequence $(u(\hbar))_{\hbar\in S}$. 
A priori, several limit measures $\mu$ may be extracted from the original
sequence, corresponding to different subsequences $S$. 
Semiclassical measures were investigated in the case of closed chaotic system (say, the geodesic
flow on a compact Riemannian manifold of negative curvature). Any
semiclassical measure associated with the eigenstates of $P(\hbar)=-\hbar^2\Delta_X/2$
must be invariant w.r.t. the classical flow. Furthermore, the quantum ergodicity
theorem states that, as long as the flow is
ergodic w.r.t. the Liouville measure, then
one can extract a subsequence $S$ of density one\footnote{This means that
this sequence contains ``almost all'' the eigenstates.}, such that
$(W^\hbar_{u(\hbar)})_{\hbar\in S}$ converges to the Liouville measure on $p^{-1}(1/2)$ \cite{Schn74,Zeld87,CdV85}. 

In the frameworks of potential scattering \cite{NZ2} or open chaotic
maps \cite{Keating+06,NoRu07}, a generalization of the above
invariance property was obtained for semiclassical measures associated
with sequences of metastable states.
\begin{thm}\label{thm:decay}\cite{NZ2}Consider a scattering Hamiltonian
  \eqref{e:Hamiltonian-V} such that for some $E>0$ the
  trapped set $K_E$ is a hyperbolic repeller. Take a sequence of resonances
$(z(\hbar)\in D(E,C\hbar))_{\hbar\to 0}$, and extract a subsequence
$(W^\hbar_{u(\hbar)})_{\hbar\in S}$ converging to a semiclassical measure $\mu$ on
$T^*B(0,R_0)$. 

Then $\mu$ will be invariant up to a decay rate $\Lambda\geq 0$:  
\be\label{e:decay}
\forall t\geq 0,\quad \Phi^{t*}\mu = e^{-t\Lambda}\,\mu\qquad \text{inside the interaction region
}T^* B(0,R_0)\,.
\ee
Furthermore, the subsequence $S$ must be such that the resonances
$(z(\hbar))_{\hbar\in S}$ satisfy
\be\label{e:z-Lambda}
\lim_{\hbar\in S,\hbar\to 0} \frac{\Im z(\hbar)}{\hbar} = -\frac{\Lambda}{2}\,.
\ee
\end{thm}
Any measure satisfying \eqref{e:decay} (at least inside the interation
region) will be called a $\Lambda$-eigenmeasure for the flow. In the
case of an open map, a $\Lambda$-eigenmeasure is characterized by the property
\be\label{e:decay1}
\kappa^*\mu = e^{-\Lambda}\,\mu\,.
\ee
$\Lambda$-eigenmeasures are easy to classify. For instance, in the
case of open maps, each $\Lambda$-eigenmeasure is uniquely determined by its
restriction on $\cK^+\cap \kappa(V)\setminus V$, which can be 
arbitrary. All such
measures are supported on $\cK^+$, and satisfy
$\mu(\cK)=0$ for $\Lambda>0$, while they are supported on $\cK$ iff
$\Lambda=0$. 

The above theorem shows that any semiclassical
measure is necessarily a $\Lambda$-eigenmeasure, with decay rate
$\Lambda$ equal to the asymptotic quantum decay rates. 
In view of the quantum ergodicity result for chaotic closed systems,
the following question naturally arises:

Given $\Lambda\geq 0$, and considering a sequence of resonances $(z(\hbar))_{\hbar\in S}$ satisfying
\eqref{e:z-Lambda}, {\it which} $\Lambda$-eigenmeasures can be obtained as
semiclassical measures? Is
there a "favoured" limit, or even a unique one? 

This question presumes that there exist sequences of
resonances satisfying \eqref{e:z-Lambda}, a fact which depends on the
semiclassical distribution of resonances; in case the strong form
\eqref{e:FWL-strong} of fractal Weyl law holds, such a sequence
exists if the profile function satisfies $\frac{dF}{d\gamma}(\Lambda/2)>0$.

We have noticed before that, according to several numerical results,  the density of resonances often shows a
peak near the value $\Lambda=\gamma_{cl}$. For this specific
value of $\Lambda$, there exists a ``natural'' $\Lambda$-measure,
which is obtained by iterating an initial smooth measure $\mu_0$ (with
support intersecting $\cK^-$):
$$
\mu_{nat} = \lim_{t\to\infty} \mathcal{N}_t\,\Phi^{t*}\mu_0\,,\quad
\text{respectively}\quad \mu_{nat} = \lim_{n\to\infty} \mathcal{N}_n\,\kappa^{n*}\mu_0\,,
$$
with $\mathcal{N}_t$, $\mathcal{N}_n$ appropriate normalization
factors (see \eqref{e:class-decay}). Yet, the
study of \cite{NoRu07} did not reveal that this measure played any
particular role for the open quantum baker's map.

In \cite{Keating+06} the
authors computed 
averages of the spatial
densities $|u_j(x)|^2$ over a few eigenstates with comparable decay rates, for the
symmetric open $3$-baker. They noticed
strong self-similar properties of the densities, depending on the
decay rates. Some of the individual Husimi functions of
\cite{NoRu07} were also featuring
a selfsimilar behaviour in both the momentum and position directions.

Rigorous results were obtained in the case of
the Walsh-quantized open baker's map \cite{Keating+08,NoRu07}, using explicit formulas for the
eigenstates. In this model most eigenvalues $\lambda_j(\hbar)$ have
large multiplicities, leaving a lot of freedom to construct
eigenstates.  
In \cite{NoRu07,Keating+08} it was shown that, for the Walsh-quantized
symmetric $3$-baker, any semiclassical sequence of eigenstates
$(u(\hbar))_{\hbar\to 0}$ with eigenvalues converging towards the
outer circle $|\lambda(\hbar)|\to r_{\max}$ (resp. the inner circle
$|\lambda(\hbar)|\to r_{\min}$) of the nontrivial spectrum, converges
to a \emph{single}
semiclassical measure $\mu_{\max}$ (resp. $\mu_{\min}$), which is of
Bernoulli type, therefore perfectly selfsimilar. This is a form of
``quantum unique ergodicity'' at the edges of the nontrivial spectrum.
On the opposite, for any value $r\in
(r_{\min},r_{\max})$, we exhibited many semiclassical measures
associated with sequences $(u(\hbar))_{\hbar\to 0}$ of asymptotic
decay rates $|\lambda(\hbar)|\to r$. For $r=e^{-\gamma_{cl}/2}$ we showed
that the natural measure $\mu_{nat}$ is \emph{not} a semiclassical measure.

\subsection{Scattering states\label{s:scatt}}

Metastable states appear in expansions of the resolvent of
$P(\hbar)$, and consequently in expansions for the time dynamics
\cite{TanZw00,BuZw01}. Another class of generalized eigenstates is
more natural from the point of view of scattering theory, namely the
scattering states, used to define the scattering matrix (see \S\ref{s:resonances}). In the
semiclassical setting of a scattering Hamiltonian $P(\hbar)$ on a manifold $X$, a scattering
state at energy $E>0$ is a wavefunction $u_E=u_E(\hbar)$ satisfying the
differential equation $(P(\hbar)-E)u_E=0$, and satisfying certain conditions at
infinity. 

If $X\equiv \IR^d$ outside the interaction region $B(0,R_0)$, one can expand
$u_E(x)$ using a basis of incoming and outgoing waves, as in Eqs.~(\ref{e:in-out},\ref{e:u_in}). 
Fixing the incoming part of $u_{E,in}$ near infinity uniquely
determines the full wavefunction $u_E$, and in particular determines
its outgoing part $u_{E,out}$,
the relation between $u_{E,in}$ and $u_{E,out}$ defining
the scattering matrix $S(E)$.
We ask the following question:

Given $u_{E,in}$, what is the spatial (or phase space) structure of
$u_E$ inside the interaction region? 

In the semiclassical/high-frequency limit, the usual basis states for
the incoming wave $u_{E,in}$ (namely the angular momentum eigenstates,
see \eqref{e:u_in})
are Lagrangian states associated with  certain Lagrangian submanifolds of the energy
shell, for instance a spherically symmetric incoming wave sits on the Lagrangian
manifold $\{(x,\xi=-\sqrt{2E}x/|x|)\}$. 
Most of the trajectories on this manifold will be scattered inside the interaction region and
then exit it towards infinity after a short transient
time. Still, a small fraction of the incoming trajectories
may be trapped during a long time in this region, travelling close
close to $K_E$, or even be trapped for ever if they exactly belong to the incoming tail
$K_E^-$. How do these trapped (or long transient) trajectories
influence the structure of $u_E$?
\begin{figure}[ht]
\begin{center}
\includegraphics[angle=00,width=1.\textwidth]{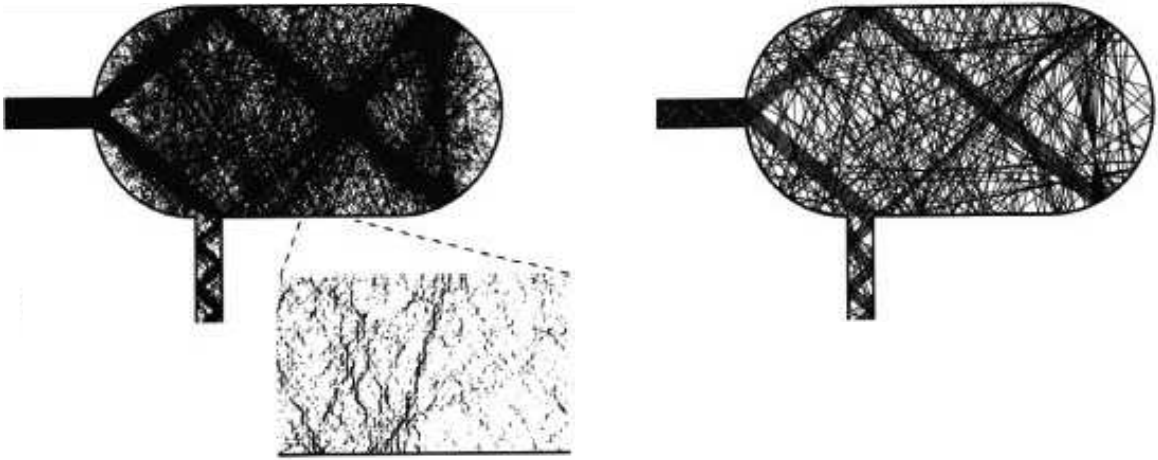}
\caption{Left: density plot of a scattering state $|u_E(x)|^2$, with
  incoming part $u_{E,in}$ a plane wave in the left lead into the
  stadium-shaped cavity. Right: corresponding classical trajectories incoming from the left opening with angles $\pm\theta_n$.
  Reprinted with permission from
  \href{http://stacks.iop.org/JPhysA/37/L217}{H. Ishio and
    J.P. Keating, 
J. Phys. {\bf A 37}
  (2004) L217--L223}. Copyright
  2004 by the Institute of Physics. \label{f:IshioKeat}}
\end{center}
\end{figure}

This question has been studied numerically by Ishio and Keating
\cite{IshKeat04} in a different geometry, namely the case of a 2d
chaotic cavity opened by two
infinite ``leads'' (waveguides). In this case, the incoming wave
$u_{E,in}$ is given by plane waves inside the left lead, $u_E(x,y)=\sin(k_ny)e^{ i k_l x}$, where the
longitudinal and transverse wavevectors $k_l,k_n$ satisfy
$$
k_n = \pi n/L,\quad E=\frac{\hbar^2}{2}(k_l^2 + k_n^2)\,,\quad
n\in\IN\setminus 0\,,\qquad \text{$L$ the width of the lead.}
$$
This incoming wave semiclassically corresponds to a pencil of trajectories coming out of
the lead with an angle $\pm\theta_n$, $\theta_n = \arcsin
(k_n/k_l)$. Two scattering states with such incoming components were numerically computed in
\cite{IshKeat04}, one of them is shown on the left of Fig.~\ref{f:IshioKeat}. In both cases the
density $|u_E(x)|^2$ is strongly imprinted by
short transient orbits. The authors also derived an
approximate semiclassical expression for $u_E(x)$, as a sum over
classical trajectories, and showed that this expression is quite
accurate for the two examples of states they have computed. They distinguished between two complementary
situations: for {\it weakly open} situations, the
contributions of long trajectories is important (even divergent); on the opposite, for {\it very open systems},
the contribution of long trajectories decays exponentially fast, so
that the wavefunction is mainly influenced by the short transient
trajectories. This dichotomy is of course reminiscent of the one
mentioned in \S\ref{s:questions}.

More recently, Guillarmou and Naud \cite{GuiNaud11} studied the scattering states for
convex co-compact manifolds $X=\Gamma\setminus\IH^{n+1}$, also called
the Eisenstein functions in this context (the spectral parameter is $s=n/2+it$, $t\gg 1$). 
A convenient ``basis'' consists in the functions $u_{s,y}(x)$ which become singular
when $x$ converges to a given point $y$ of the boundary $\partial X$:
the incoming wave is then associated with the unstable manifold made of the
geodesics issued from $y$. In such a homogeneous situation, the
wavefunction $u_{s,y}(x)$ can be simply expressed by a sum over the group
$\Gamma$. The authors are able to precisely describe
$u_{s,y}$ provided the trapped set is ``thin'', that is the dimension of the
limit set satisfies $\delta < n/2$ (equivalently, $\cP(-\varphi^+/2)<0$). This
is a precise criterion for the ``very open'' situation of \cite{IshKeat04}.
One can then compute the semiclassical measure associated
with the family $(u_{n/2+it,y})_{t\to\infty}$: it is an invariant
measure supported by the full unstable
manifold issued from $y$. 


If one averages the densities $|u_{s,y}(x)|^2$ over
the boundary point $y$, one recovers the uniform (Haar) measure on $X$,
plus a semiclassically small correction given by a sum over periodic orbits,
similarly with Gutzwiller's trace formula for closed systems (one
difference being that the sum over the orbits is absolutely
convergent). 

This description of scattering states can certainly be extended to more
general geometries or systems with a ``thin'' hyperbolic trapped set.


\section{Conclusion}\label{s:concl}
We have presented several analytical methods used to analyze the
spectral properties of scattering operators in the semiclassical/high
frequency limit, in cases where the set of classically trapped trajectories
is a hyperbolic repeller. In particular, the number of long-living resonances near some classical
energy $E>0$ was bounded from above by a fractal power of the
semiclassical parameter, reflecting the fact that these long living
states must be supported on the trapped set, which is a fractal subset
of the energy shell. We stated two types of ``fractal Weyl law'' conjectures,
predicting that this upper bound should be sharp, and presented some
numerical results in favor of these laws, both for
scattering flows and for the model of open quantum maps. 

A second result is the presence of a ``resonance gap'' (or
a uniform lower bound for the quantum decay rates), provided the instability of the
flow exceeds its complexity (precisely, provided the topological
pressure $\cP(-\varphi^+/2)$ is negative). This criterion allows
to split such chaotic scattering systems between ``very open''
vs. ``weakly open'' systems. We showed that this dichotomy was also 
relevant in the precise description of scattering wavefunctions. 

At the technical level, we presented
quantum monodromy operators associated with a quantum scattering flow,
which can be used
to investigate this spectral problem. These operators, which contain
the full long living quantum dynamics, can be deformed such as
to live in a ``minimal'' neighbourhood of the trapped set, still faithfully
representing the ``quantum mechanics on the trapped set''.
They resemble Ruelle transfer operators appearing in
classical dynamics. Hopefully, a more precise analysis of these
operators could deliver some nontrivial information on the resonance
spectrum, like a proof of the fractal Weyl law (under some genericity
assumption) or a sharper criterion for a resonance free strip.

The resonances were analyzed as eigenvalues of certain
nonselfadjoint pseudodifferential operators. The techniques presented
above can also be used in a different context, namely the study of a
``closed'' quantized chaotic systems in the presence of some
``damping'', e.g. the case of damped waves propagating on
a manifold of negative curvature. In that case there is no  ``escape to
infinity'', but the high frequency spectral
problem presents
similar features \cite{Sj00}. For instance, fractal Weyl upper bounds were
obtained for such systems \cite{Anan10}, and a bound for the decay
rates in terms of a topological pressure  was also proved in this context
\cite{Schenck10,Schenck11}, with applications to the stabilization
of the damped waves. 

The same type of ideas could also be useful when describing the scattering by a
{\it dielectric cavity}, relevant in the description
of quasi-2d microlasers (see e.g. \cite{WieMai08,Shino+09,Bogo+11} and
references therein). In such situations, the damping is due to the fact
that a wavepacket propagating inside the cavity loses a fraction of its
energy when being reflected by the boundary of the cavity, the
rest of the energy being refracted outside to infinity. How does the
shape of the cavity influence the resonance spectrum, in particular in
case the internal dynamics is chaotic? How do the
metastable states look like? To my
knowledge, the rigorous studies of such cavities have so far been restricted to strictly convex cavities
with smooth boundaries \cite{CarPopVod01}, for which the ray dynamics
cannot be purely chaotic. 

\appendix

\section{A brief review of $\hbar$-pseudodifferential
  calculus}\label{s:appen}

We recall some definition and basic properties of
Weyl's quantization, in the semiclassical setting. 
For simplicity, we will only consider operators on the Euclidean space
$\IR^d$.  See \cite[Chapter 7]{DiSj} for a detailed discussion of 
semiclassical quantization,
\cite[Chap.4, Part 3]{EZB} for the pseudodifferential calculus for
the symbol classes presented below, and \cite[Chap.13]{EZB}
for its generalization to the calculus on manifolds.

\subsection{Weyl quantization and pseudodifferential calculus}\label{s:Weyl}
Weyl's quantization associates to a smooth phase space function $a\in
C^\infty(T^*\IR^d) $  (the {\it classical observable}, or {\it symbol}) an operator 
acting on $u\in C^\infty_c(\RR^d)$ as follows:
\be\label{eq:weyl}
\begin{split}
[a^w\, u](x) & 
=  [\Op (a) u ] ( x) \\
& \defeq \frac1{ ( 2 \pi \hbar )^d } 
  \int \int  a \Big( \frac{x + y  }{2}  , \xi \Big) 
e^{ i \la x -  y, \xi \ra / \hbar } u ( y ) dy d \xi \,.
\end{split} 
\ee
In these notations, $\hbar\in (0,1]$ is Planck's ``constant'' (which
we always assume to be ``small''). 
The integral converges absolutely only if $a(x,\xi)$ decays fast
enough w.r.t. $\xi$, but by integrating by parts one
can easily extend the definition to functions growing algebraically in
$\xi$. The classes of symbols presented below are engineered such that
the above formula makes sense.

Weyl's quantization leads to the definition of the Wigner
distribution $W^\hbar_u$ associated with a function $u\in L^2$, by the
following duality:
\be\label{e:Wigner}
\forall a\in C^\infty_c(T^*\IR^d),\qquad \la W^\hbar_u,a\ra_{(C^\infty_c)',C^\infty_c} \defeq \la
u,\Op(a) u\ra_{L^2}
\ee

When $\hbar$ is small,
the product of two operators $\Op(a)\Op(b)$ can be analyzed through their
symbols $a$, $b$. That product is itself an operator
of the form $\Op(c)$, with a symbol $c(x,\xi)$ given by the Moyal
product of $a$ and $b$:
\be\label{e:expansion}\begin{split}
c &= a\sharp_\hbar b \defeq a\, \exp\left(\frac{i h}{2} (\la
  \overleftarrow{D}_\xi , \overrightarrow{D}_x \ra - \la
  \overleftarrow{D}_x,\overrightarrow{D}_\xi \ra)  \right)\, b \\
& \sim a\,b +  \frac{i\hbar}{2}
\{a, b\} + \sum_{j\geq 2} \frac{(i\hbar/2)^j}{j!}\, 
a \,\Big(\la\overleftarrow{D}_\xi , \overrightarrow{D}_x \ra 
- \la\overleftarrow{D}_x,\overrightarrow{D}_\xi\ra \Big)^j \, b\,,
\end{split}\ee
where $D_\bullet=-i\partial_\bullet$, and $\{a,b\}$ is the Poisson
bracket. The above sum  is a good asymptotic expansion when $\hbar\to
0$, in the sense that the sum up to the term $j=N-1$
gives a good approximation of $c$,  with a
remainder $\cO(\hbar^N)$. It is at the heart of {\it
  pseudodifferential calculus}.
Even if $a,b$ are independent of $\hbar$, the symbol $c$ does depend on $\hbar$. It
thus makes sense to define classes of $\hbar$-dependent symbols, characterized by the
regularity property of $a(x,\xi;\hbar)$, uniformly in the limit $\hbar\to 0$.

One standard class of symbols is the following: for $k\in \IR$, let 
\be\begin{split} 
S^{k} ( T^* \RR^d ) =\Big\{ & a \in C^\infty( T^* \RR^d_{x,\xi} \times (0, 1]_{\hbar}
  ) :\ \forall \alpha,\beta\in\IN^d,\\
& \ |\partial_x ^{ \alpha } \partial _\xi^\beta a ( x, \xi ;\hbar ) | \leq
C_{\alpha,\beta} 
(1+ |\xi|)^{k-|\beta| } \Big\}  \,, 
\end{split}
\ee
The improved decay in $\xi$ upon differentiation
is necessary for the class to
be invariant upon a smooth change of coordinates, which is
crucial when extending the formalism to manifolds.
The corresponding operator classes are
denoted by $ \Psi^{k} ( \RR^d) $. For instance, the Schr\"odinger
operator \eqref{e:Hamiltonian-V} is the Weyl
quantization of the symbol $p(x,\xi)=\frac{|\xi|^2}{2}+V(x)\in S^2(T^*\IR^d)$.

These symbol classes are closed under
composition: for $a\in S^{k}$, $b\in
S^{\ell}$, the product operator $\Op(a)\Op(b)=\Op(c)$ belongs to $\Psi^{k+\ell}$.
An important property  is the action on
$L^2(\IR^d)$. For $a\in
S^0(T^*\IR^d)$, the operators $\Op(a)$ are bounded on $L^2(\IR^d)$,
with 
\be\label{e:C-V}
\| \Op(a)\|_{L^2\to L^2} = \|a(\hbar)\|_{L^\infty} + \cO(\hbar)\,.
\ee
If $a(x,\xi)$ is real valued, $\Op(a)$ will be self-adjoint on
$L^2$. 
In this case, one can also analyze functions of $\Op(a)$ using
their symbols: for a smooth function $f:\IR\to\IR$, the
operator $f(\Op(a))$ belongs to $\Psi^0(\IR^d)$, with symbol
$f(a)+\cO(\hbar)$.
For instance, in \S\ref{s:escape1}  the operators $e^{\pm tG^w}$
belong to $\Psi^0(\IR^d)$, and the composition rule \eqref{e:expansion} shows
that the conjugated operator
$e^{-t G^w}\,P_\theta(\hbar)\,e^{t G^w}$ belongs to $\Psi^2(\IR^d)$,
with a symbol of the form \eqref{e:symbol-p-thetaG}.

\subsection{Exotic symbol classes}\label{s:exotic}
The symbols $a\in S^0(T^*\IR^d)$ fluctuate on distances $\sim
1$. For our purposes, we also needed to consider symbols fluctuating on microscopic distances.
For $k\in\IR$ and $\delta\in [0,1/2]$, we consider the ``exotic''
symbol classes
\be\begin{split} 
S^{k}_\delta ( T^* \RR^d ) =\Big\{ & a \in C^\infty( T^* \RR^d \times (0, 1]
  ) :\ \forall \alpha,\beta\in\IN^d,\\
& \ |\partial_x ^{ \alpha } \partial _\xi^\beta a ( x, \xi ;h ) | \leq
C_{\alpha,\beta} \,\hbar^{-\delta ( | \alpha| + |\beta |) } \,
(1+|\xi|)^{k-|\beta| } \Big\}  \,,
\end{split}\ee
which encompasses symbols
fluctuating on distances $\gtrsim \hbar^\delta$. 
For $\delta<1/2$, the expansion
\eqref{e:expansion} makes sense, and we can still use the symbol
to analyze the operator. 
These exotic classes were used to construct
the exponential weights of \S\ref{s:escape1}. If we take
$\vareps=\hbar^\delta$, the escape
function $G(x,\xi)$  must belong to 
the class $\log(1/\hbar) S^0_{\delta}(T^*X)$ (see the model function
$G_1(x,\xi)$ of \eqref{e:G_1}), and the corresponding
functional calculus allows to
analyze the operators $e^{\pm tG^w}$ and
$e^{-tG^w}P(\hbar)e^{tG^w}\in \Psi^2_{\delta}(\IR^d)$. The symbol
$\alpha$ of \eqref{e:alpha} also belongs to an exotic class $S^0_\delta$. 

\subsection{Fourier integral operators}\label{s:FIO}
A time dependent Hamiltonian $p(t,x,\xi)\in
C([0,1]_t,S^2(T^*\IR^d))$ generates a nonautonomous symplectic flow
$(\kappa_t)_{t\in[0,1]}$ through Hamilton's equations
$$
\frac{d\kappa_t}{dt}=(\kappa_t)_* H_{p(t)}\,,\quad \kappa_0=Id,\quad
t\in [0,1]\,.
$$
Then, the family of unitary operators $U(t)$ defined by 
$$
i\hbar \partial_t U(t) = U(t)\, p^w(t),\quad U(0)=Id\,,
$$
defines a family of quantum propagators, which are
unitary Fourier Integral Operators (FIO) associated with the
diffeomorphisms $\kappa_t$. 

Consider the propagator $U=U(1)$ associated with $\kappa=\kappa_1$. 
$U$ maps a wavepacket microlocalized at $(x_0,\xi_0)$ to a wavepacket
localized at $\kappa(x_0,\xi_0)$. Its action on a quantum
observable satisfies a quantum-classical correspondence (called Egorov's
theorem in the mathematical literature): for any symbol $a\in
S^0(T^*\IR^d)$ of compact support, one
has
\be\label{e:Egorov0}
U^{-1}\,\Op(a)\,U = \Op(b),\quad b\in S^0(T^*\IR^d),\quad b=a\circ \kappa + \cO(\hbar)\,.
\ee
More generally, an FIO associated with $\kappa$ will be an operator of
the form 
\be
\cM(\alpha,\hbar)= U\,\Op(\alpha)\,,
\ee
with $\alpha\in S^0_\delta(T^*\IR^d)$ for some
$\delta\in [0,1/2)$. The FIOs of \S\ref{s:OQM}, in particular the open quantum maps, are of
this type. From there one easily shows the ``nonunitary'' Egorov
property \eqref{e:Egorov}. Also, the $L^2$ norm estimate \eqref{e:norm-est}
is obtained from \eqref{e:C-V}.

\end{document}